\documentclass[11pt,letter]{article}
\newtheorem{theorem}{Theorem}[section]
\newtheorem{definition}{Definition}[section]

\newtheorem{lemma}{Lemma}[section]
\newtheorem{proposition}{Proposition}[section]

\usepackage{amsmath}
\usepackage{amssymb}

\begin{document}

\title{Adaptive Concurrent    Non-Malleability    \\ with Bare
Public-Keys\thanks{The work described in this paper was supported
in part by the National Basic Research Program of China Grant
(973) 2007CB807900, 2007CB807901, and  by a grant from the
Research Grants Council of the Hong Kong Special Administrative
Region, China (Project Number CityU 122105) and CityU Research
Grant (9380039).
 The third author is also supported by 
 NSFC (No. 60703091), the Pu-Jiang and Qi-Ming-Xing
Programs of Shanghai and a grant from MSRA.} }


\author{Andrew  C. Yao\footnote{Institute for Theoretical Computer Science  (ITCS), Tsinghua University, Beijing, China.
 \texttt{andrewcyao@tsinghua.eud.cn} } 
\and Moti Yung\footnote{Google Inc.  and Columbia University, New
York, NY, USA. \texttt{moti@cs.columbia.edu}}
 \and Yunlei Zhao\footnote{Contact author. Software School,   Fudan University, Shanghai 200433,
 China.    \texttt{ylzhao@fudan.edu.cn} Works partially done while visiting Tsinghua university and
City University of Hong Kong. }}
\date{}          
\maketitle
\renewcommand\thefootnote{\fnsymbol{footnote}} 

\renewcommand{\thefootnote}{\arabic{footnote}} 

\begin{abstract}
Concurrent non-malleability (CNM) is central  
for cryptographic protocols running concurrently in environments
such as the Internet. 
 In this work, we formulate CNM in the bare
public-key (BPK) model, and show that round-efficient concurrent
non-malleable cryptography with \emph{full} adaptive input
selection can be established, in general,  with bare public-keys
(where, in particular, no trusted assumption is made). Along the
way, we clarify the various 
 subtleties of adaptive
concurrent non-malleability in the bare public-key  model.

\end{abstract}

\newpage

\tableofcontents

\newpage



\section{Introduction}

Concurrent non-malleability  is  central  for  cryptographic
protocols secure against concurrent man-in-the-middle  (CMIM)
attacks. In
the CMIM setting, 
 polynomially many
concurrent executing instances  (sessions)  of a  protocol
  take place in an asynchronous setting
(appropriate for environments such as over the Internet),  and all
the unauthenticated communication channels (among all the
concurrent sessions) are controlled by a probabilistic
polynomial-time (PPT) CMIM adversary
$\mathcal{A}$.  
 In this setting, honest players are assumed
oblivious of each other's existence, nor do they generally know
the topology of the network, and thus cannot coordinate their
executions. 
The CMIM adversary $\mathcal{A}$ (controlling the communication
channels) can
do whatever it wishes. 
When CNM with adaptive input selection is considered,
$\mathcal{A}$
can also set input to each   session. 


 Unfortunately,  in the stringent CMIM setting, large
classes of cryptographic functionalities cannot be securely
implemented round-efficiently, and even cannot be securely
implemented  with \emph{non-constant} round-complexity against
\emph{adaptive input selecting} CMIM adversaries in the plain
model \cite{CKPR01,Ljoc,L03focs}. In such cases, some setup
assumptions are necessary, and establishing the general
feasibility of round-efficient concurrent non-malleable
cryptography with adaptive input selection,  with setups as
minimal as possible,  has been being a basic problem extracting
intensive research efforts in the literature.

In this work, we investigate CNM security in the bare public-key
model (introduced by Canetti, Goldreich, Goldwasser and Micali
\cite{CGGM00}).
 A protocol in the BPK model  simply assumes that all  players  have each deposited a public key in a
public file before any interaction takes place among the users.
Note that, no assumption is made on whether the public-keys
deposited are unique or valid (i.e., public keys can even be
``nonsensical," where no corresponding secret-keys exist or are
known) \cite{CGGM00}. That is, no trusted third party is assumed,
the underlying communication network is assumed to be
adversarially
asynchronous, and 
preprocessing is reduced to minimally non-interactively posting
public-keys in a public file.  
%
%
 In many cryptographic settings, availability of a public key
infrastructure (PKI) is assumed or required and in these settings
the BPK model is, both, natural and attractive (note that the BPK
model is, in fact, a weaker version of PKI where in the later
added key certification is assumed).  It was pointed out by Micali
and Reyzin \cite{MR01a} that BPK is, in fact, applicable to
interactive systems in general.

\subsection{Our contributions}
 We examine   concurrent non-malleability in the BPK
model, by investigating two types of protocols,  specifically,
zero-knowledge (ZK) \cite{GMR89} and coin-tossing (CT) \cite{B82}
both of which are central and fundamental to modern cryptography.

We show the insufficiency of existing CNM formulations in the
public-key model,  reformulate CNM zero-knowledge (CNMZK) and CNM
coin-tossing (CNMCT) in the BPK model. 
   The CNMCT definition implies (or serves as a general basis to formulate) the CNM security for  
 for any cryptographic protocol in the BPK model against
\textsf{CMIM with full adaptive input selection}. By \textsf{CMIM
with \emph{full} adaptive input selection}, we mean that  the CMIM
adversary can set inputs to all concurrent  sessions; furthermore
and different from the  traditional formulation of adaptive input
selection, the adversary does not necessarily  set the input to
each session at the beginning of the session; Rather, the input
may be set on the way of the session, and is based on the whole
transcript evolution (among other concurrent sessions and the
current session). Also, 
 similar to \cite{YYZ07}, we allow the CMIM adversary to adaptively set the language to be proved in right sessions,  based on players' public-keys and  common statements of left sessions. 
We motivate the desirability of achieving CNM security against
\textsf{CMIM with full adaptive input selection}, and  clarify the
various   subtleties of  the CNM formulations and make in-depth
discussions. The CNM reformulations,  and the various subtlety
clarifications and discussions, constitute independent
contributions
of this work, 
 which provide the insight for understanding the 
complex and subtle nature of (adaptive)  CNM with bare public-keys.

We then  present a constant-round CNMCT protocol 
 in the BPK model under standard
assumptions, which is enabled by the recent celebrated Pass-Rosen
ZK (PRZK) result \cite{PR05s,PR05}. The importance of the CNMCT
protocol is that it can be used to transform concurrent
non-malleable protocols that are originally developed in the
common random string (CRS) model into the weaker BPK model (with full adaptive input
selection). That
is, round-efficient
 concurrent non-malleable cryptography (with full adaptive input
selection)  can be established with  bare public-keys,  in
general.


\subsection{Related works} \label{relatedworks}

The concept of non-malleability is introduced by Dolve, Dwork and
Naor in the seminal work of  \cite{DDN91}. The work of
\cite{DDN91} also presents non-constant-round non-malleable
commitment and zero-knowledge protocols. Constant-round
non-malleable coin-tossing protocol in the plain model (and
accordingly, constant-round non-malleable zero-knowledge arguments
for $\mathcal{NP}$ and commitment schemes by combining the result
of \cite{DDOPS01})  is
 achieved by Barak  \cite{B02}. The non-malleable
coin-tossing protocol of \cite{B02} employs non-black-box
techniques (introduced in \cite{B01}) in a critical way. CNMZK with a poly-logarithmic  round complexity is achieved in the plain model \cite{BPS06}.

A large number of concurrent non-malleable (and the strongest,
universal composable) cryptographic protocols are developed in the
common reference/random string model, where a common
reference/random  string is selected trustily by  a trusted third
party and is known to all players (e.g.,
\cite{DIO98,FF00,DKOS01,DDOPS01,CLOS02,DG03}, etc). The work of
\cite{KL05} demonstrates the general feasibility of concurrent
non-malleability with timing assumption, where each party has a
local clock and all clocks proceed at approximately the same rate.

The CNMCT formulation and construction presented in this work are
based on the incomplete work of \cite{Z03}, but with significant
extension and correction in view of the recent advances of
security formulations (e.g.,  the formulation of secret-key
independent knowledge-extraction \cite{YYZ07}) and  non-malleable
building tools
(e.g., the PRZK result \cite{PR05s,PR05}).  
The situation with  adaptive concurrent non-malleability in the
bare public-key model turns out to be notoriously subtle and
somewhat confused. There are several works that deal with
concurrent non-malleability
 in the BPK model \cite{Z03,OPV06,DDL06,OPV07} (the works of \cite{OPV06,DDL06,OPV07} consider
 the specific protocol, specifically CNMZK, in the BPK model). But, a careful
investigation shows that the CNM formulations in all existing
works are flawed or incomplete (details can be found in Section
\ref{CNMrevisited}). Also, 
 no previous protocols in the BPK model or the plain model  can be proved CNM secure
against \textsf{CMIM with full adaptive input selection}.
Actually, the possibility of CNM with adaptive input selection in
the BPK model  itself turns out to be a subtle issue, and was not
clarified in existing works.

\section{Preliminaries}\label{preliminaries}

\textbf{Basic notation.}  We use standard notations and
conventions below for writing probabilistic algorithms,
experiments and interactive protocols. If \emph{A} is
 a probabilistic algorithm, then $A(x_1, x_2, \cdots; r)$ is the result of running \emph{A} on inputs
 $x_1, x_2, \cdots$ and coins $r$. We let $y\leftarrow A(x_1, x_2, \cdots)$ denote the experiment of picking $r$ at
 random and letting $y$ be $A(x_1, x_2, \cdots; r)$. If $S$ is a finite set then $x\leftarrow S$ is the operation of
 picking an element uniformly from $S$. If $\alpha$ is neither an algorithm nor a set then $x\leftarrow \alpha$ is a
 simple assignment statement.  By $[R_1; \cdots; R_n: v]$ we
 denote the set of values of $v$ that a random variable can
 assume, due to the distribution determined by the sequence of
 random processes $R_1, R_2, \cdots, R_n$. By $\Pr[R_1; \cdots; R_n: E]$ we denote
 the probability of event $E$, after the ordered execution of
 random processes $R_1, \cdots, R_n$.

 Let $\langle P, V\rangle$ be a
 probabilistic
 interactive protocol, then the notation $(y_1, y_2)\leftarrow
 \langle P(x_1), V(x_2)\rangle (x)$ denotes the random process of
 running interactive protocol $\langle P, V \rangle$ on common input $x$, where $P$
 has private input $x_1$, $V$ has private input $x_2$, $y_1$ is
 $P$'s output and $y_2$ is $V$'s output. We assume w.l.o.g. that the
 output of both parties $P$ and $V$ at the end of an execution of
 the protocol $\langle P, V \rangle$ contains a transcript of the
 communication exchanged between $P$ and $V$ during such
 execution.

The security  of cryptographic primitives and tools,  presented
throughout this work,  is defined with respect to uniform
polynomial-time algorithms (equivalently, polynomial-size
circuits). When it comes to non-uniform security, we refer to
non-uniform polynomial-time algorithms (equivalently, families of
polynomial-size circuits).

On a security parameter $n$ (also written as $1^n$), a function
$\mu(\cdot)$ is \textsf{negligible} if for every polynomial
$p(\cdot)$, there exists a value $N$ such that for all $n>N$ it
holds that $\mu(n)<1/p(n)$. Let $X=\{X(n, z)\}_{n\in N, z\in \{0,
1\}^*}$ and $Y=\{Y(n, z)\}_{n\in N, z\in \{0, 1\}^*}$ be
distribution ensembles. Then we say that $X$ and $Y$ are
\textsf{computationally (resp., statistically) indistinguishable},
if for every probabilistic polynomial-time (resp., any, even power-unbounded) algorithm $D$, 
 for all sufficiently large $n$'s, and every $z\in \{0, 1\}^*$, $|\Pr[D(n, z,
X(n, z))=1]-\Pr[D(n, z, Y(n, z))=1]|$ is negligible in $n$.

\begin{definition}[one-way
function]\label{OWF} A function $f: \{0, 1\}^*\longrightarrow \{0,
1\}^*$ is called a  one-way function (OWF) if the following
conditions hold:
\begin{enumerate}
\item Easy to compute: There exists a (deterministic)
polynomial-time algorithm $A$ such that on input $x$ algorithm $A$
outputs $f(x)$ (i.e., $A(x)=f(x)$).

\item Hard to invert: For every probabilistic polynomial-time PPT
algorithm $A^{\prime}$, every positive polynomial $p(\cdot)$, and
all sufficiently large $n$'s, it holds $\Pr[A^{\prime}(f(U_n),
1^n)\in f^{-1}(f(U_n))]<\frac{1}{p(n)}$, where $U_n$ denotes a
random variable uniformly distributed over $\{0, 1\}^n$.

\end{enumerate}

\end{definition}

\begin{definition}[interactive argument/proof system]
A pair of  interactive  machines, $\langle P,\,V \rangle$, is
called an interactive argument  system for a language
$\mathcal{L}$ if both are probabilistic polynomial-time (PPT)
machines and  the following conditions hold:
\begin{itemize}
\item Completeness. For every $x\in \mathcal{L}$, there exists a
string $w$ such that for every string $z$,  \\ $\Pr[\langle
P(w),\,V(z) \rangle(x)=1]=1$.

\item Soundness. For every polynomial-time interactive machine
$P^*$, and for all sufficiently large $n$'s and every $x\notin
\mathcal{L}$ of length $n$ and every $w$ and $z$, $\Pr[\langle
P^*(w),\,V(z) \rangle(x)=1]$ is negligible in $n$.
\end{itemize}
An interactive protocol is called a \emph{proof} for
$\mathcal{L}$, if the soundness condition holds against any (even
power-unbounded) $P^*$ (rather than only PPT $P^*$). An
interactive  system is called a public-coin  system if at each
round the prescribed verifier can only toss coins and send their
outcome  to the prover.
\end{definition}

\begin{definition}[witness indistinguishability WI \cite{FS90}]\label{definitionWI}
Let $\langle P,\,V \rangle$ be an interactive system for a
language $\mathcal{L}\in \mathcal{NP}$, and let
$\mathcal{R}_\mathcal{L}$ be the fixed $\mathcal{NP}$ witness
relation for $\mathcal{L}$. That is, $x \in \mathcal{L}$ if there
exists a $w$ such that $(x,\,w)\in \mathcal{R}_\mathcal{L}$. We
denote by $view_{V^*(z)}^{P(w)}(x)$ a random variable describing
the transcript of all messages exchanged between a (possibly
malicious) PPT verifier $V^*$ and the honest prover $P$ in an
execution of the protocol on common input $x$, when $P$ has
auxiliary input $w$ and $V^*$ has auxiliary input $z$. We say that
$\langle P,\,V \rangle$ is witness indistinguishable for
$\mathcal{R}_{\mathcal{L}}$ if for every PPT interactive machine
$V^*$, and every two sequences $W^1=\{w^1_x\}_{x\in L}$ and
$W^2=\{w^2_x\}_{x\in L}$ for sufficiently long $x$, so that
$(x,\,w^1_x)\in \mathcal{R}_{\mathcal{L}}$ and $(x,\, w^2_x)\in
\mathcal{R}_{\mathcal{L}}$, the following two probability
distributions are computationally indistinguishable by any
non-uniform polynomial-time algorithm:
$\{x,\,view_{V^*(z)}^{P(w^1_x)}(x)\}_{x\in \mathcal{L},\, z\in
\{0,\,1\}^*}$ and $\{x,\,view_{V^*(z)}^{P(w^2_x)}(x)\}_{x\in
\mathcal{L},\, z\in \{0,\,1\}^*}$. Namely, for every  non-uniform
polynomial-time distinguishing algorithm $D$, every polynomial
$p(\cdot)$, all sufficiently long $x\in \mathcal{L}$, and all
$z\in \{0, 1\}^*$, it holds that
$$|\Pr[D(x, z, view_{V^*(z)}^{P(w^1_x)}(x)=1]-\Pr[D(x, z,
view_{V^*(z)}^{P(w^2_x)}(x)=1]|<\frac{1}{p(|x|)}$$

\end{definition}
It is interesting to note that the  WI property preserves  against
adaptive concurrent composition \cite{FS90,F90,FLS99,DDOPS01}.

\begin{definition}[strong witness indistinguishability SWI \cite{G01}]
Let $\langle P, V\rangle$ and all other notations be as in
Definition \ref{definitionWI}. We say that $\langle P, V\rangle$
is \emph{strongly witness-indistinguishable for
$\mathcal{R}_{\mathcal{L}}$} if for every PPT interactive machine
$V^*$ and for every two probability ensembles $\{X^1_n, Y^1_n,
Z^1_n\}_{n\in N}$ and $\{X^2_n, Y^2_n, Z^2_n\}_{n\in N}$, such
that each $\{X^i_n, Y^i_n, Z^i_n\}_{n\in N}$ ranges over
$(\mathcal{R}_\mathcal{L}\times \{0, 1\}^*)\cap (\{0, 1\}^n \times
\{0, 1\}^* \times \{0, 1\}^*)$, the following holds: If $\{X^1_n,
Z^1_n\}_{n\in N}$ and $\{X^2_n, Z^2_n\}_{n\in N}$ are
computationally indistinguishable, then so are $\{\langle
P(Y^1_n), V^*(Z^1_n)\rangle (X^1_n)\}_{n\in N}$ and $\{\langle
P(Y^2_n), V^*(Z^2_n)\rangle (X^2_n)\}_{n\in N}$.

\end{definition}

\textbf{WI vs. SWI:} It is clarified in \cite{G02f} that the
notion of SWI actually refers to issues that are fundamentally
different from WI. Specifically, the issue is whether the
interaction with the prover helps $V^*$ to distinguish some
auxiliary information (which is indistinguishable without such an
interaction). Significantly different from WI, SWI does \emph{not}
preserve under concurrent composition. More details about SWI are
referred to \cite{G02f}. An interesting observation, as clarified
later,  is: the protocol composing commitments and SWI can be
itself regular WI. Also note that any zero-knowledge protocol is
itself SWI \cite{G02f}.

\begin{definition} [zero-knowledge ZK \cite{GMR89,G01}] Let $\langle P,\,V\rangle$ be an interactive
system for a language $\mathcal{L}\in \mathcal{NP}$, and let
$\mathcal{R}_{\mathcal{L}}$ be the fixed $\mathcal{NP}$ witness
relation for $\mathcal{L}$. That is, $x\in \mathcal{L}$ if there
exists a $w$ such that $(x,\,w)\in \mathcal{R}_\mathcal{L}$. We
denote by $view_{V^*(z)}^{P(w)}(x)$ a random variable describing
the contents of the random tape of $V^*$ and the messages $V^*$
receives from $P$ during  an execution of the protocol on common
input $x$, when $P$ has auxiliary input $w$ and $V^*$ has
auxiliary input $z$.
 Then we say that $\langle P,\,V\rangle$ is  zero-knowledge
if for  every probabilistic polynomial-time interactive machine
$V^*$  there exists a probabilistic (expected) polynomial-time
oracle machine $S$, such that for all sufficiently long $x\in
\mathcal{L}$ the ensembles $\{view_{V^*}^{P(w)}(x)\}_{x \in
\mathcal{L}}$ and $\{S^{V^*}(x)\}_{x \in \mathcal{L}}$
 are computationally indistinguishable. Machine $S$ is called a ZK simulator for
$\langle P,\,V\rangle$. The protocol is called statistical ZK if
the above two ensembles are statistically close (i.e., the
variation distance is eventually smaller than $\frac{1}{p(|x|)}$
for any positive polynomial $p$). The protocol is called perfect
ZK if the above two ensembles are actually  identical (i.e.,
except for
 negligible probabilities, the two ensembles are
equal).
\end{definition}

\begin{definition}[system for argument/proof of knowledge \cite{G01,BG06}]
\label{POK} Let $\mathcal{R}$ be a binary relation and $\kappa:
N\rightarrow [0, 1]$. We say that  a probabilistic polynomial-time
(PPT) interactive machine $V$ is a knowledge verifier for the
relation $\mathcal{R}$ with knowledge error $\kappa$ if the
following two conditions hold:
\begin{itemize}
\item Non-triviality: There exists an interactive machine $P$ such
that for every $(x, w)\in \mathcal{R}$ all possible interactions
of $V$ with $P$ on common input $x$ and auxiliary input $w$ are
accepting. \item Validity (with error $\kappa$): There exists a
polynomial $q(\cdot)$ and a probabilistic oracle machine $K$ such
that for every interactive machine $P^*$, every $x\in
\mathcal{L}_\mathcal{R}$, and every $w, r\in\{0, 1\}^*$, machine
$K$ satisfies the following condition:

Denote by $p(x, w, r)$ the probability that the interactive
machine $V$ accepts, on input $x$, when interacting with the
prover specified by $P^*_{x, w, r}$ (where $P^*_{x, w, r}$ denotes
the strategy of $P^*$ on common input $x$, auxiliary input $w$ and
random-tape $r$). If $p(x, w, r)>\kappa(|x|)$, then, on input $x$
and with oracle access to $P^*_{x, w, r}$, machine $K$ outputs a
solution $w^{\prime}\in \mathcal{R}(x)$ within an expected number
of steps bounded by
$$\frac{q(|x|)}{p(x, w, r)-\kappa (|x|)}$$ The oracle machine $K$ is
called a knowledge extractor.

\end{itemize}
An interactive argument/proof system $\langle P, V\rangle$  such
that $V$ is a knowledge verifier for a relation $\mathcal{R}$ and
$P$ is a machine satisfying the non-triviality condition (with
respect to $V$ and $\mathcal{R}$) is called a system for
argument/proof of knowledge (AOK/POK) for the relation
$\mathcal{R}$.
\end{definition}

The above definition of POK is with respect to
\emph{deterministic} prover strategy. POK also can be defined with
respect to \emph{probabilistic} prover strategy. It is recently
shown that the two definitions are equivalent for all natural
cases (e.g., POK for $\mathcal{NP}$-relations) \cite{BG06}.

\begin{definition}[pseudorandom functions PRF]
On a security parameter $n$, let $d(\cdot)$ and $r(\cdot)$ be two
positive polynomials in $n$. We say that $$\{f_s: \{0,
1\}^{d(n)}\longrightarrow \{0, 1\}^{r(n)}\}_{s\in \{0, 1\}^n}$$ is
a
 pseudorandom function ensemble  if the following two conditions
 hold:
 \begin{enumerate}
 \item Efficient evaluation: There exists a polynomial-time
 algorithm that on input $s$ and $x\in \{0, 1\}^{d(|s|)}$ returns
 $f_s(x)$.
 \item Pseudorandomness: For every probabilistic polynomial-time
 oracle machine $A$, every polynomial $p(\cdot)$, and all
 sufficiently large $n$'s, it  holds:  $$|\Pr[A^{F_n}(1^n)=1]-\Pr[A^{H_n}(1^n)=1]|<\frac{1}{p(n)}$$
where $F_n$ is a random variable uniformly distributed over the
multi-set $\{f_s\}_{s\in\{0, 1\}^n}$, and $H_n$ is  uniformly
distributed among all functions mapping $d(n)$-bit-long strings to
$r(n)$-bit-long strings.
\end{enumerate}
\end{definition}

 PRFs can be constructed under
any one-way function \cite{GGM86,G01}. The current most practical
 PRFs are the Naor-Reingold implementations under the factoring (Blum integers) or the decisional
 Diffie-Hellman  hardness assumptions \cite{NR04}. The computational complexity of computing the
 value of the Naor-Reingold functions at a given point is  about
 two modular exponentiations and can be further reduced to only
 two multiple products modulo a prime (without any exponentiations!) with natural preprocessing,
 which is great for practices involving PRFs.


\begin{definition} [statistically/perfectly binding bit commitment scheme]
A  pair of PPT interactive machines,
 $\langle P, V\rangle$, is called a perfectly binding bit commitment scheme, if it satisfies  the following:

\begin{description}
\item [Completeness.] For any security parameter $n$,  and any bit
$b\in \{0, 1\}$, it holds that \\ $\Pr[(\alpha, \beta)\leftarrow
\langle P(b), V\rangle (1^n); (t, (t, v))\leftarrow \langle
P(\alpha), V(\beta)\rangle(1^n): v=b]=1$.

\item [Computationally hiding.]  For all sufficiently large $n$'s,
any PPT adversary $V^*$, the following two probability
distributions are  computationally indistinguishable: $[(\alpha,
\beta)\leftarrow \langle P(0), V^*\rangle (1^n): \beta]$  and
$[(\alpha^{\prime}, \beta^{\prime})\leftarrow \langle P(1),
V^*\rangle (1^n):
\beta^{\prime}]$. 

\item [Perfectly Binding.] For all sufficiently large $n$'s, and
\emph{any}  adversary $P^*$, the following probability is
negligible (or equals 0 for perfectly-binding commitments):
$\Pr[(\alpha, \beta)\leftarrow \langle P^*, V\rangle (1^n); (t,
(t, v))\leftarrow \langle P^*(\alpha), V(\beta)\rangle(1^n);
(t^{\prime}, (t^{\prime}, v^{\prime}))\leftarrow \langle
P^*(\alpha), V(\beta)\rangle(1^n): v, v^{\prime} \in \{0, 1\}
\bigwedge v\neq v^{\prime}]$.

That is, no (\emph{even computational power unbounded}) adversary
$P^*$ can decommit the same transcript of the commitment stage
both to 0 and 1.

\end{description}

\end{definition}

Below, we recall some classic perfectly-binding commitment
schemes.

 One-round
perfectly-binding (computationally-hiding) commitments can be
  based on any one-way permutation OWP \cite{B82,GMW91}.
Loosely speaking, given a OWP $f$ with a hard-core predict $b$
(cf. \cite{G01}), on a security parameter $n$ one commits  a bit
$\sigma$ by uniformly selecting $x\in \{0, 1\}^n$ and sending
$(f(x), b(x)\oplus \sigma)$ as a commitment, while keeping $x$ as
the decommitment information.


Statistically-binding commitments can be  based on any one-way
function (OWF) but run in two rounds \cite{N91,HILL99}. On a
security parameter $n$, let $PRG: \{0, 1\}^n\longrightarrow \{0,
1\}^{3n}$ be a pseudorandom generator, the Naor's OWF-based
two-round public-coin perfectly-binding commitment scheme works as
follows: In the first round, the commitment receiver sends a
random string $R\in \{0, 1\}^{3n}$ to the committer. In the second
round, the committer uniformly selects a string $s\in \{0, 1\}^n$
at first; then to commit  a bit 0 the committer sends $PRG(s)$ as
the commitment; to commit  a bit 1 the committer sends
$PRG(s)\oplus R$ as the commitment. Note that the first-round
message of Naor's commitment scheme can be fixed once and for all
and, in particular, can be posted as a part of public-key in the
public-key model.

\textbf{Commit-then-SWI:} Consider the following protocol
composing
  a statistically-binding commitment and SWI:

  \begin{description}
  \item [Common input:] $x\in \mathcal{L}$ for an $\mathcal{NP}$-language $\mathcal{L}$
  with corresponding $\mathcal{NP}$-relation $\mathcal{R}_{\mathcal{L}}$.
  \item [Prover auxiliary input:] $w$ such that $(x, w)\in \mathcal{R}_\mathcal{L}$.
  \item[The protocol:] consisting of two stages:
  \begin{description}
  \item [Stage-1:] The prover $P$ computes and sends $c_w=C(w,
  r_w)$, where $C$ is a statistically-binding commitment and $r_w$
  is the randomness used for commitment.

  \item [Stage-2:] Define a new language $\mathcal{L}^{\prime}=\{(x, c_w)|
   \exists (w, r_w) \ s.t.\ c_w=C(w, r_w)
  \wedge \mathcal{R}_\mathcal{L}(x, w)=1  \}$. Then, $P$ proves to $V$ that it knows a
  witness to $(x, c_w)\in \mathcal{L}^{\prime}$, by running a SWI protocol for $\mathcal{NP}$.

  \end{description}
  \end{description}

  One interesting observation for the above commit-then-SWI protocol
  is that commit-then-SWI is itself a regular WI for $\mathcal{L}$.

  \begin{proposition} \label{CSWI}  Commit-then-SWI is itself a regular WI for
  the language $\mathcal{L}$.\end{proposition}

  \textbf{Proof} (of Proposition \ref{CSWI}). For any PPT malicious
  verifier   $V^*$, possessing some auxiliary input $z\in \{0, 1\}^*$,
  and for any $x\in \mathcal{L}$ and two (possibly different) witnesses $(w_0,
  w_1)$ such that $(x, w_b)\in \mathcal{R}_\mathcal{L}$ for both $b\in \{0, 1\}$,
  consider the executions of commit-then-SWI: $\langle P(w_0), V^*(z)\rangle (x)$ and
$\langle P(w_1), V^*(z)\rangle (x)$.

Note that for $\langle P(w_b), V^*(z)\rangle (x)$, $b\in \{0,
1\}$, the input to SWI of Stage-2 is $(x, c_{w_b}=C(w_b,
r_{w_b}))$, and the auxiliary input to $V^*$  at the beginning of
Stage-2 is $(x, c_{w_b}, z)$. Note that $(x, c_{w_0}, z)$ is
indistinguishable from $(x, c_{w_1}, z)$. Then, the regular WI
property of the whole composed  protocol is followed from the SWI
property of Stage-2. \hfill $\square$

\subsection{Adaptive tag-based one-left-many-right non-malleable statistical
zero-knowledge argument of knowledge (SZKAOK)} \label{appPRZK}

Let $\{\langle P_{TAG}, V_{TAG}\rangle (1^n) \}_{n\in N, TAG\in
\{0, 1\}^{p(n)}}$, where $p(\cdot)$ is some polynomial, be a
family of argument systems for an $\mathcal{NP}$-language
$\mathcal{L}$ specified by $\mathcal{NP}$-relation
$\mathcal{R}_{\mathcal{L}}$. For each security parameter $n$ and
$TAG\in \{0, 1\}^{p(n)}$, $\langle P_{TAG}, V_{TAG}\rangle (1^n)$
is an instance of the protocol $\langle P, V\rangle$, which is
indexed by $TAG$ and  works for inputs in $\mathcal{L}\cup \{0,
1\}^n$.

 We consider
an experiment $\textup{EXPE}(1^n, x, TAG, z)$, where $1^n$ is the
security parameter, $x\in \mathcal{L}\cup \{0, 1\}^n$,  $TAG\in
\{0, 1\}^{p(n)}$ and $z\in \{0, 1\}^*$. (The input $(x, TAG)$
captures the predetermined input and tag of the prover instance in
the following left MIM part, and the string $z\in \{0, 1\}^*$
captures the auxiliary input to the following MIM adversary
$\mathcal{A}$.) In the experiment $\textup{EXPE}(1^n, x, TAG, z)$, on
input $(1^n, x, TAG, z)$, an adaptive input-selecting
one-left-many-right MIM
adversary $\mathcal{A}$ 
 is simultaneously participating in two
interaction parts:  

\begin{description}
\item [The left MIM part:] in which $\mathcal{A}$ chooses
$(\tilde{x}^l, \widetilde{TAG}^l)$
 based on its view from both the left session and all right
sessions, satisfying that: the membership of $\tilde{x}^l\in
\mathcal{L}\cup \{0, 1\}^n$ can be efficiently checked
\footnote{We remark, for our purpose of security analysis in
Section \ref{analysis}, it is necessary, as well as  sufficient,
to require the membership of the statement $\tilde{x}^l$ chosen by
$\mathcal{A}$ can be efficiently checked; otherwise, the
experiment may render an $\mathcal{NP}$-membership oracle to
$\mathcal{A}$.} and $\widetilde{TAG}^l\in \{0, 1\}^{p(n)}$; In
case $\tilde{x}^l\in \mathcal{L}\cup \{0, 1\}^n$ (that can be
efficiently checked), then a witness $\tilde{w}^l$ such that
$(\tilde{x}^l, \tilde{w}^l)\in \mathcal{R}_\mathcal{L}$ is given
to the prover instance $P_{\widetilde{TAG}^l}$, and  $\mathcal{A}$
interacts, playing the role of the verifier
$V_{\widetilde{TAG}^l}$, with the prover instance
$P_{\widetilde{TAG}^l}(\tilde{x}, \tilde{w})$ on common input
$\tilde{x}^l$ . The interactions with
$P_{\widetilde{TAG}^l}(\tilde{x}^l, \tilde{w}^l)$ is called the
left session. Note that, $\mathcal{A}$ can just set $(\tilde{x}^l,
\widetilde{TAG}^l)$ to be $(x, TAG)$, which captures the case of
predetermined input and tag to left session.

\item [The right CMIM part:]  in which $\mathcal{A}$
\emph{concurrently} interacts with $s(n)$, for a polynomial
$s(\cdot)$, verifier instances:
$V_{\widetilde{TAG}^r_1}(\tilde{x}^r_1)$,
$V_{\widetilde{TAG}^r_2}(\tilde{x}^r_2)$, $\cdots$,
$V_{\widetilde{TAG}^r_{s(n)}}(\tilde{x}^r_{s(n)})$, where
$(\widetilde{TAG}^r_i, \tilde{x}^r_i)$, $1\leq i\leq s(n)$, are
set by $\mathcal{A}$ (at the beginning of each session) adaptively
based on its view (in both the left session and all the right
sessions) satisfying $\tilde{x}^r_i \in \{0, 1\}^n$ and
$\widetilde{TAG}^r_i\in \{0, 1\}^{p(n)}$. The interactions with
the instance $V_{\widetilde{TAG}^r_i}(\tilde{x}^r_i)$ is called
the $i$-th right session, in which $\mathcal{A}$ plays the role of
$P_{\widetilde{TAG}^r_i}$.


\end{description}

Denote by $view_{\mathcal{A}}(1^n, x, TAG, z)$ the random variable
describing the view of $\mathcal{A}$ in the above experiment
$\textup{EXPE}(1^n, x, TAG, z)$, which includes the input $(1^n,
x, TAG, z)$, its random tape, and all messages received in the one
left session and the $s(n)$ right sessions.

Then, we say that  the family of argument systems $\{\langle
P_{TAG}, V_{TAG}\rangle (1^n) \}_{n\in N, TAG\in \{0, 1\}^{p(n)}}$
is \textsf{adaptive tag-based one-left-many-right non-malleable
SZKAOK with respect to tags of length $p(n)$},  if for any PPT
adaptive input-selecting one-left-many-right MIM adversary
$\mathcal{A}$ defined above, there exists an expected
polynomial-time algorithm $S$, such that for any sufficiently
large $n$, any $x\in \mathcal{L}\cup \{0, 1\}^n$ and $TAG\in \{0,
1\}^{p(n)}$, and any $z\in \{0, 1\}^*$,  the output of $S(1^n, x,
TAG, z)$ consists of two parts $(str, sta)$ such that the
following hold, where we denote by $S_1(1^n, x, TAG,
z)$ (the distribution of)  its first output $str$. 

\begin{itemize}
\item \textbf{Statistical simulatability.} The following ensembles
are statistically indistinguishable: \\ $\{view_{\mathcal{A}}(1^n,
x, TAG, z)\}_{n\in N,x\in \mathcal{L}\cup \{0, 1\}^n, TAG\in \{0,
1\}^{p(n)}, z\in \{0, 1\}^*}$ and \\ $\{S_1(1^n, x, TAG,
z)\}_{n\in N,x\in \mathcal{L}\cup \{0, 1\}^n, TAG\in \{0,
1\}^{p(n)}, z\in \{0, 1\}^*}$

\item \textbf{Knowledge extraction.} $sta$ consists of a set of
$s(n)$ strings, $\{w_1, w_2, \cdots, w_{s(n)}\}$, satisfying the
following:
\begin{itemize}
\item For any $i$, $1\leq i\leq s(n)$, if the $i$-th right session
in $str$ is aborted or with a tag identical to that of the left
session, then $w_i=\bot$; \item Otherwise, i.e., the $i$-th right
session in $str$ is successful with  $\widetilde{TAG}^r_i\neq
\widetilde{TAG}^l$, then $(\tilde{x}^r_i, w_i)\in
\mathcal{R}_\mathcal{L}$, where $\tilde{x}^r_i$ is the input to
the $i$-th right session in $str$.
\end{itemize}

\end{itemize}

\textbf{Pass-Rosen ZK (PRZK).} The PRZK developed in
\cite{PR05s,PR05} is the only known  \emph{constant-round}
adaptive tag-based one-left-many-right non-malleable  SZKAOK, that
is based on any collision-resistant hash function (that can in turn be based on the existence of a family of claw-free permutations). Furthermore,
PRZK is public-coin.


\section{The CMIM Setting in the BPK Model with \emph{Full}  Adaptive Input Selection}\label{CMIMBPK}
In this section, we  clarify the
subtleties of adaptive input selection in the CMIM setting, and
motivate the desirability for  CNM security against CMIM
adversaries of the capability of  \emph{full} adaptive input
selection. Then,  we describe the CMIM setting in the BPK model in
accordance with any interactive argument protocol (that works for a class of admissible languages rather than a unique language). 

\subsection{Motivation for CMIM with \emph{full} adaptive input
selection}\label{fulladaptive}


 A concurrent man-in-the-middle (CMIM) adversary $\mathcal{A}$, for an interactive proof/argument protocol,
 is a probabilistic polynomial-time (PPT) algorithm
that can act both as a prover and as a verifier. Specifically,  $\mathcal{A}$ can
concurrently interact with any polynomial number of instances of
the honest prover  in \emph{left interaction
part}. The interactions with each instance of the honest prover  is called a \emph{left session}, in which $\mathcal{A}$
plays the role of the verifier;
\emph{Simultaneously}, $\mathcal{A}$ interacts with any polynomial
number of instances of the  honest verifier  in \emph{right
interaction part}. The interactions with each instance of the
honest verifier  is called a \emph{right session}, in which
it plays the role of the  prover.
Here, all honest prover and verifier instances are working
independently, and answer messages sent by $\mathcal{A}$ promptly.

 In the traditional formulation of the CMIM settings (and also the stand-alone MIM settings),
 there are two levels of input-selecting capabilities
for the CMIM adversary: (1) \textsf{CMIM  with predetermined
left-session inputs}, in which the inputs to left sessions are
predetermined, and the CMIM adversary $\mathcal{A}$ can only set
inputs to right sessions; (2) \textsf{CMIM with adaptive input
selection}, in which $\mathcal{A}$ can set, adaptively based on
its view, the inputs to both left sessions and right sessions.
But, in the  traditional formulation of CMIM, both for
\textsf{CMIM with predetermined left-session inputs} and for
\textsf{CMIM with adaptive input selection}, the CMIM adversary
$\mathcal{A}$ is required (limited) to set the input of each
session at the beginning of that   session. We note that this
requirement, on input selection in traditional
CMIM formulation,  could essentially limit 
 the power
of the CMIM adversary in certain natural settings. 
We give some concrete examples below.

 Consider any protocol resulted from the composition of a
coin-tossing protocol and a protocol in the CRS model. In most
often cases,  the input to the underlying protocol in the CRS
model, denoted CRS-protocol for notation simplicity,  is also the
input to the whole composed protocol. Note that the input to the
underlying CRS-protocol   can be set
 after the coin-tossing phase is finished, furthermore, can be
set only at the last message of the composed protocol. We remark
that it is true that for adaptive adversary in the CRS model, it
is allowed to set statements based on the CRS. In other words,
mandating the adversary to predetermine the input to the
underlying CRS-protocol, without seeing the output of coin-tossing
that serves as the underlying CRS, clearly limits the power of the
adversary and thus weakens  the provable security established for
the composed
protocol. 

Another example is the Feige-Shamir-ZK-like protocol
\cite{FS89,F90,Z03}, which consists of two sub-protocols (for
presentation convenience, we call them verifier's sub-protocol and
prover's sub-protocol) and the input of the protocol is only used
in the prover's sub-protocol. The prover can set and prove the
statements in the prover's sub-protocol, only after the verifier
has successfully finished the verifier's sub-protocol in which the
verifier proves some knowledge (e.g., its secret-key) to the
prover. In this case, the adversary can take advantage of the
verifier's sub-protocol interactions  to set and prove inputs to
the subsequent prover's sub-protocol,
 especially when the Feige-Shamir-ZK-like protocol  is run \emph{concurrently}  in the
 \emph{public-key} model.
Again, an adversary, as well as the honest prover, could set the
input to a session only at the last message of the session, for
example, considering  the  prover's sub-protocol is  the
Lapidot-Shamir WIPOK protocol \cite{LS90}.  As demonstrated in
\cite{YZ06, YZ07, YYZ07} and in  this work, letting the adversary
adaptively determine inputs, in view of the concurrent executions
of the verifier's sub-protocol in the public-key model,
  renders strictly stronger power to the adversary.

In contrast, by \textsf{CMIM with \emph{full} adaptive input
selection}, we mean that  a CMIM adversary can set inputs to both
left sessions and right sessions; furthermore (and different from
the  traditional formulation of adaptive input selection), the
adversary does not necessarily  set the input to each session at
the beginning of the session; Rather, the input may be set on the
way of the session, and is based on the whole transcript evolution
(among other concurrent sessions and the current session); Though
the adversary is allowed to set inputs at any points of the
concurrent execution evolution, whenever at some point the
subsequent activities of an honest player in a session may utilize
the input of the session while the adversary did not provide the
input, the honest player just simply aborts the session. Similar
to traditional \textsf{CMIM with predetermined left-session
inputs}, we can define \textsf{CMIM with predetermined
left-session inputs but full adaptive input selection on the
right}, in which the inputs to left sessions are fixed and the
CMIM adversary only sets inputs to right session in the above
fully adaptive way.


  From above clarifications, we conclude that allowing the CMIM adversary the capability
  of full adaptive input selection, in particular  not necessarily predetermining the inputs of sessions
   at the start
of each session, is a more natural formulation, as well as more
natural scenarios,  for  cryptographic protocols to be CNM-secure
against adaptive input selecting CMIM adversaries.
It renders stronger capability to the adversary, and thus allows
us to achieve stronger provable CNM security. The general CNM
feasibility in the BPK model established in this work is against
CMIM with the capability of full adaptive input selection (and the capability of adaptive  language selection for right sessions).


\subsection{The CMIM Setting in the BPK Model (with adaptive input and language selection)}\label{CMIMBPK}

\textbf{The  bare public-key (BPK) model.} As in \cite{YYZ07}, we say a class of $\mathcal{NP}$-languages $\mathcal{L}$ is \emph{admissible} to a protocol
$\langle P, V\rangle$ if the protocol can work (or, be
instantiated) for any language $L\in \mathcal{L}$. Typically,
$\mathcal{L}$ could be the set of all $\mathcal{NP}$-languages
 or the set of any
languages admitting $\Sigma$-protocols (in the latter  case $\langle P,
V\rangle$ could be instantiated for any language in $\mathcal{L}$
efficiently without going through general
$\mathcal{NP}$-reductions).  We assume that given the description of the  corresponding $\mathcal{NP}$-relation $\mathcal{R}_L$ of an $\mathcal{NP}$-language $L$, the admissibility of $L$ (i.e., the membership of  $L\in \mathcal{L}$) can be efficiently decided.

Let
$\mathcal{R}^P_{KEY}$ be an $\mathcal{NP}$-relation validating the
public-key and secret-key pair $(PK_P, SK_P)$ generated by honest
provers, i.e., $\mathcal{R}^P_{KEY}(PK_P, SK_P)=1$ indicates that
$SK_P$ is a valid secret-key of $PK_P$. Similarly, let
$\mathcal{R}^V_{KEY}$  be an $\mathcal{NP}$-relation validating
the public-key and secret-key pair $(PK_V, SK_V)$ generated by
honest verifiers, i.e., $\mathcal{R}^V_{KEY}(PK_V, SK_V)=1$
indicates that $SK_V$ is a valid secret-key of $PK_V$. In the
following formalization, we assume each honest player is of fixed
player role.

Then, a protocol $\langle P, V\rangle$ for an
$\mathcal{NP}$-language $L$ in the BPK model w.r.t.
key-validating relations $\mathcal{R}^P_{KEY}$ and
$\mathcal{R}^V_{KEY}$, consists of the following:


\begin{enumerate}
\item The  interactions between $P$ and $V$  can be divided into
two stages. The first stage is called \emph{key-generation stage}
in which each player registers a public-key in a public file $F$;
at the end of the key-generation stage, the \emph{proof stage}
starts, where any pair of prover and verifier can interact. All
algorithms have access to the same  public file $F$ output by the
key-generation stage.

\item On security parameter $1^n$, the \emph{public file} $F$,
structured as a collection of $poly(n)$ records, for a polynomial
$poly(\cdot)$:  $\{(id_1, PK_{id_1}), (id_2, PK_{id_2}), \cdots
(id_{poly(n)}, PK_{id_{poly(n)}})\}$. $F$ is empty at the
beginning  and is updated by players during  the key-generation
stage. As we assume players be of fixed roles, for presentation
simplicity, we also denote $F=\{PK^{(1)}_I, PK^{(2)}_I, \cdots,
PK^{(poly(n))}_I\}$, such that for any $i$, $1\leq i \leq
poly(n)$, $PK^{(i)}_I$ denotes a prover-key if $I=P$ or a
verifier-key if $I=V$.
 The \emph{same}  version of the public file $F$ obtained at the end of the
key-generation stage will be used during the proof stage. That is,
the public file $F$ to be used in proof stages remains intact with
that output at the end of key-generation stage.


\item An honest \emph{prover} $P$ is a pair of deterministic
polynomial-time algorithm $(P_1,  P_2)$, where $P_1$ operates in
the key-generation stage and $P_2$ operates in the proof stage. On
input a security parameter $1^n$ and a random tape $r_{P_1}$,
$P_1$ generates  a key pair $(PK_P, SK_P)$ satisfying
$\mathcal{R}^P_{KEY}(PK_P, SK_P)=1$, registers $PK_P$ in the
public file $F$ as its public-key while keeping the corresponding
secret key $SK_P$ in secret. Denote by $\mathcal{K}_P$ the set of
all legitimate (in accordance with $\mathcal{R}^P_{KEY}$)
public-keys generated by $P_1(1^n)$, that is, $\mathcal{K}_P$
contains all possible legitimate prover public-key generated on
security parameter $n$. Then, in the proof stage, on inputs
$(PK_P, SK_P)$, and $poly(n)$-bit string $x\in L$, an
auxiliary input $w$, a public file $F$ and a verifier public-key
$PK^{(j)}_V \in F$, and a random tape $r_P$, $P_2$ performs an
interactive protocol with
 the verifier of $PK^{(j)}_V$ in the proof
stage.

\item An honest \emph{verifier} $V$ is a pair of deterministic
polynomial-time algorithm $(V_1,  V_2)$, where $V_1$ operates in
the key-generation stage and $V_2$ operates in the proof stage. On
input a security parameter $1^n$ and a random tape $r_{V_1}$,
$V_1$ generates  a key pair $(PK_V, SK_V)$ satisfying
$\mathcal{R}^V_{KEY}(PK_V, SK_V)=1$, registers $PK_V$ in the
public file $F$ as its public-key while keeping the corresponding
secret key $SK_V$ in secret.  Denote by $\mathcal{K}_V$ the set of
all legitimate (in accordance with $\mathcal{R}^V_{KEY}$)
public-keys generated by $V_1(1^n)$, that is, $\mathcal{K}_V$
contains all possible legitimate verifier public-key generated on
security parameter $n$. On inputs $(PK_V, SK_V)$, the public file
$F$ and a prover public-key $PK^{(j)}_P \in F$, the $\mathcal{NP}$-relation $\mathcal{R}_L$, and a
$poly(n)$-bit $x$ and a random tape $r_{V_2}$,  $V_2$ first checks the admissibility of  $L\in \mathcal{L}$; Then, $V$  performs the
interactive protocol with (the proof stage of) the prover of
$PK^{(j)}_P$, and outputs ``accept $x\in L$" or ``reject $x$" at the
end of this protocol. \emph{We stress that as the role of the
honest verifier with its public-key is not interchangeable in the
BPK model, the honest verifier  with its public-key may prove the
knowledge of its secret-key, but  will never prove anything
else.} 


\end{enumerate}

\textbf{Notes:} We remark that, though each player is allowed to
register public-keys in the public-file in the original
formulation of the BPK model \cite{CGGM00}, for some cryptographic
tasks, e.g., concurrent and resettable zero-knowledge, only
requiring verifiers to register public-keys suffices. In  these
cases  provers' keys may not be  used, or $\mathcal{K}_P$ can be
just empty. Our formulation of the BPK model is for the general
case, and provers' registered public-keys play an essential role
for achieving CNM security with \emph{full} adaptive input
selection (to be addressed later). Also note that in the above
formulation, honest players are of fixed roles. For protocols with
players of interchangeable roles, the direct extension approach is
to let each player register a pair of public-keys $(PK_P, PK_V)$
and explicitly indicate its role in protocol executions.

\textbf{The CMIM adversary.}  The CMIM adversary $\mathcal{A}$ in
the  BPK model is a probabilistic polynomial-time (PPT) algorithm
that can act both as a prover and as a verifier, both in the
key-generation stage and in the main proof stage.

In the key-generation stage,  on $1^n$ and some auxiliary input
$z\in \{0, 1\}^*$ and a pair of honestly generated public-keys
$(PK_P, PK_V)$ generated by the honest prover and verifier,
$\mathcal{A}$ outputs a set of public-keys, denoted by
$F^{\prime}$, together with some auxiliary information $\tau$ to
be used in the proof-stage (in particular $\tau$ can include $z$
and  a priori information about the secret-keys of honest players
$(SK_P, SK_V)$). Then the public file $F$ used in proof state is
set to be $F^{\prime}\cup\{PK_P, PK_V\}$. That is, $\mathcal{A}$
has complete control of the public file $F$. Here, we remark that,
in general, the input to $\mathcal{A}$ in order to generate
$F^{\prime}$ could be a set of public-keys generated by many
honest provers and verifiers, rather than a single pair of
public-keys $(PK_P, PK_V)$ generated by  a single honest prover
and a single honest verifier. The formulation with a unique pair
of honestly generated public-keys is only for presentation
simplicity.

In the proof stage, on inputs $(F, \tau)$ $\mathcal{A}$ can
concurrently interact with any polynomial number of instances of
the honest prover of public-key $PK_P$  in \emph{left interaction
part}. The interactions with each instance of the honest prover of
$PK_P$ is called a \emph{left session}, in which $\mathcal{A}$
plays the role of verifier with a public-key $PK^{(j)}_V\in F$;
\emph{Simultaneously}, $\mathcal{A}$ interacts with any polynomial
number of instances of the  honest verifier $PK_V$  in \emph{right
interaction part}. The interactions with each instance of the
honest verifier of $PK_V$ is called a \emph{right session}, where
it plays the role of prover with a public-key $PK^{(j)}_P\in F$.
Here, all honest prover and verifier instances are working
independently, and answer messages sent by $\mathcal{A}$ promptly.

Specifically,  polynomially many concurrent sessions  of the proof
stage of the same protocol $\langle P, V \rangle$ take place in an
asynchronous setting (say, over the Internet),  and all the
\emph{unauthenticated} communication channels (among all the
concurrently executing instances of $\langle P, V\rangle$) are
controlled by the  PPT adversary $\mathcal{A}$. This means that
the honest prover instances cannot directly communicate with the
honest verifier instances in the proof stages, since all
communication messages are done through the adversary.
 The adversary $\mathcal{A}$, controlling the scheduling of messages in both parts of CMIM,
   can decide to  simply relay the messages between  any prover instance  in the left part and the corresponding verifier instance
  in the   right part. But,  it can also decide to block, delay, divert, or change messages arbitrarily at its
  wish.

  We allow the CMIM adversary to set (admissible) languages for the right sessions (possibly different from the language for the left sessions), adaptively based on all players' public-keys and the (predetermined)  statements of left sessions. Specifically, the left-sessions and right-sessions (of the same protocol)  may work for different (admissible)  languages.  For presentation simplicity, we assume the CMIM adversary sets a unique  language $\hat{L}$ (by giving the corresponding $\mathcal{NP}$-relation $\mathcal{R}_{\hat{L}}$) for all concurrent right sessions before the actual interactions of proof stages take place.
  For CMIM-adversary  with
adaptive input selection, $\mathcal{A}$ can further set the inputs
to left sessions adaptively based on its view (besides adaptively
setting inputs to right sessions).  A CMIM adversary is called
$s(n)$-CMIM adversary, for a positive polynomial $s(\cdot)$,
   if the adversary involves at most
$s(n)$ concurrent sessions in each part of the CMIM setting and
registers at most $s(n)$ public-keys in $F^{\prime}$, where $n$ is
the security parameter.  

    For presentation simplicity and without loss of generality,  we have made the following conventions:
   \begin{itemize}
   \item  We  assume  all honest prover instances are of the same public-key $PK_P$ and
       all  honest verifier instances are of
    the same public-key $PK_V$.  That is, $\mathcal{A}$
concurrently  interacts  on the left with honest prover instances
of the same public-key $PK_P$ and on the right  with honest
verifier instances of the same public-key $PK_V$. And, the file
$F^{\prime}$ generated by $\mathcal{A}$ is only based on $\{PK_P,
PK_V\}$.

\item The session number in left
   interaction part is equal to the session number in right
   interaction part, i.e., both of them are $s(n)$.

    \item We assume $\mathcal{A}$ sets
    the same $\mathcal{NP}$-relation  $\mathcal{R}_{\hat{L}}$  for all right sessions.
    \end{itemize}

    We remark that both the security model and the security analysis in this work can be easily
    extended to the  general
   case:  multiple different honest prover and verifier  instances with multiple different
   public-keys;      different session numbers
   in left interactions and right interactions;   and allowing setting
   different (admissible) languages
    for different right sessions. 
 We prefer the simplified formulation for the reason that it much simplifies
   the presentation and security analysis.    The (simplified) CMIM setting for interactive arguments  in the bare
    public-key  model
   with a PPT $s(n)$-CMIM adversary is depicted
   in Figure \ref{CMIM} (page \pageref{CMIM}).

 \vspace{0.3cm}
\begin{center}
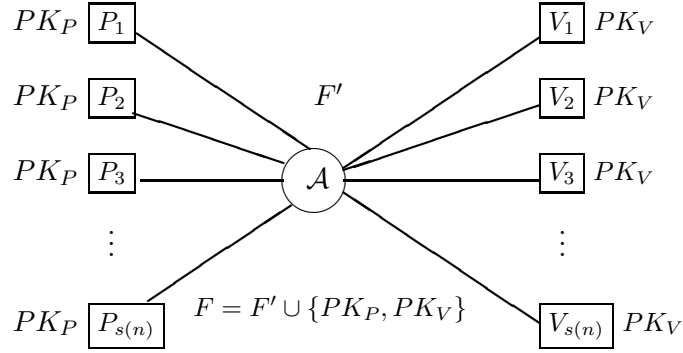
\begin{figure}[!t]
\begin{center}
\setlength{\unitlength}{1cm}
\begin{picture}(6,5)
 \put(-1,5){$PK_P$ \fbox{\small $P_1$}}
\put(-1,4){$PK_P$ \fbox{\small $P_2$}} \put(-1,3){$PK_P$
\fbox{\small $P_3$}} \put(0,2){$\ \ \vdots$} \put(-1,1){$PK_P$
\fbox{\small $P_{s(n)}$}} \put(3,4){$F^{\prime}$}
 \put(1.4, 1.2){\small $F=F^{\prime}\cup \{PK_P, PK_V\}$}
 \put(3,3){\circle{0.8}}
\put(2.9,2.9){$\mathcal{A}$}  \put(6,5){\fbox{\small $V_1$} \small
$PK_V$}
\put(6,4){\small \fbox{$V_2$} $PK_V$} \put(6,3){\small
\fbox{$V_3$} $PK_V$} \put(6,2){$\ \ \vdots$} \put(6,1){\small
\fbox{$V_{s(n)}$} $PK_V$} \thicklines
\put(0.65,4.95){\line(3,-2){2.3}} \put(0.6,3.9){\line(3,-1){2}}
\put(0.7,3){\line(3,0){1.9}} \put(0.8,1.4){\line(3,2){1.9}}
\put(6,4.9){\line(-3,-2){2.6}} \put(6,4){\line(-3,-1){2.6}}
\put(6,3){\line(-3,0){2.6}} \put(6,1.1){\line(-3,2){2.6}}

\end{picture}
 \caption{\label{CMIM} \small  The CMIM setting in the
public-key model for ZK}
\end{center}
\end{figure}
\end{center}


More formally, with respect to a protocol $\langle P, V\rangle$
for an (admissible)  $\mathcal{NP}$-language $L\in \mathcal{L}$  with
$\mathcal{NP}$-relation $\mathcal{R}_{L}$, an
$s(n)$-CMIM adversary $\mathcal{A}$'s attack in the  BPK model is
executed in accordance with the following experiment
$\textsf{Expt}^{\mathcal{A}}_{CMIM}(1^n, X, W, z)$, where
$X=\{x_1, \cdots, x_{s(n)}\}$ and $W=\{w_1, \cdots, w_{s(n)}\}$
are vectors of $s(n)$ elements such that $x_i\in L\cap \{0,
1\}^{poly(n)}$ and $(x_i, w_i)\in \mathcal{R}_{\mathcal{L}}$,
$1\leq i\leq s(n)$:

\begin{center} $\textsf{Expt}^{\mathcal{A}}_{CMIM}(1^n, X, W,  z)$ \label{CMIMexpt} \end{center}
\begin{description}
\item [Honest prover-key generation.] $(PK_P, SK_P)\longleftarrow
P_1(1^n)$. \item [Honest verifier-key generation.] $(PK_V,
SK_V)\longleftarrow V_1(1^n)$. \item [Preprocessing stage of the
CMIM.] $\mathcal{A}$, on inputs  $1^n$, auxiliary input $z\in \{0,
1\}^*$ and honest player keys $(PK_P, PK_V)$, outputs
$(F^{\prime}, \tau)$, where $F^{\prime}$ is a list of, at most
$s(n)$,  public-keys and $\tau$
is some auxiliary information 
 to be transferred to the proof stage of $\mathcal{A}$. Then,
the public file to be used in the proof stage is:  $F=F^{\prime}
\cup \{PK_P, PK_V\}$.

\item[Proof stage of the CMIM.] On $(F, \tau)$ and the predetermined left-session inputs $X$, $\mathcal{A}$ outputs  the description of the $\mathcal{NP}$-relation  $\mathcal{R}_{\hat{L}}$ for a language $\hat{L}$ (that may be different from $L$).

     Then,  $\mathcal{A}$ continues its
execution, 
 and may start
(at most) $s(n)$ sessions in either the left CMIM interaction part
or the right CMIM interaction part. At any time during this stage,
$\mathcal{A}$ can do one of the following four actions.
\begin{itemize}
\item Deliver to $V$ a message for an already started right
session.

\item Deliver to $P$ a message for an already started
left session.

\item Start a new $i$-th left session, $1\leq i\leq
s(n)$: $\mathcal{A}$ indicates a key $PK^{(j)}_V\in F$ to the
honest prover $P$ (of public-key $PK_P$). The honest prover $P$
then initiates a new session with (the predetermined) input $(x_i,
w_i)$ and the verifier of $PK^{(j)}_V$ (pretended by
$\mathcal{A}$).

For CMIM-adversary with (traditional) adaptive input selection, besides $PK^{(j)}_V$
the CMIM adversary $\mathcal{A}$ indicates to $P$,  adaptively
based
on its view,  a statement $\tilde{x}_i\in \{0, 1\}^{poly(n)}$ as the  input of as the $i$-th left session. In this case, we require that the membership of $\tilde{x}_i\in
\mathcal{L}\cup \{0, 1\}^{poly(n)}$  can be efficiently checked, otherwise, the
experiment may render an $\mathcal{NP}$-membership oracle to
$\mathcal{A}$. In
case $\tilde{x}_i\in \mathcal{L}\cup \{0, 1\}^{poly(n)}$ (that can be
efficiently checked), then a witness $\tilde{w}_i$ such that
$(\tilde{x}^l, \tilde{w}^l)\in \mathcal{R}_\mathcal{L}$ is given
to the prover instance of  $P$; Then,  on input $(x_i,
w_i)$ the honest prover $P$ interacts with the verifier of $PK^{(j)}_V$ (pretended by
$\mathcal{A}$).



\item Start a new $i$-th right session: the CMIM adversary
$\mathcal{A}$ chooses, adaptively based on its view from the CMIM
attack,
a $poly(n)$-bit string  $\hat{x}_i$,   and indicates a key
$PK^{(j)}_P\in F$ and the $\mathcal{NP}$-relation  language $\mathcal{R}_{\hat{L}}$ to the honest verifier $V$ of public-key $PK_V$;
Then, the honest verifier $V$  initiates a new session, checks the admissibility of $\mathcal{R}_{\hat{L}}$, and then
interacts  with the  prover of public-key $PK^{(j)}_P$ (pretended
by $\mathcal{A}$) on input $(1^n, \hat{x}_i, \mathcal{R}_{\hat{L}})$  in which
$\mathcal{A}$ is trying to convince of the (possibly false)
statement ``$\hat{x}_i\in
\hat{L}$".  

\item Output a special ``end attack" symbol within time polynomial
in $n$.

\end{itemize}

 We denote by $view_{\mathcal{A}}(1^n, X,   z)$ the random
variable describing the view of $\mathcal{A}$ in this experiment
$\textsf{Expt}^{\mathcal{A}}_{CMIM}(1^n, X, W, z)$, which includes
its random tape, the (predetermined) input vector $X$, the
auxiliary string $z$, all messages it receives including the
public-keys $(PK_P, PK_V)$ and all messages sent by honest prover
and verifier instances in the proof stages.   For any $(PK_P,
SK_P) \in \mathcal{R}^P_{KEY}$ and $(PK_V, SK_V)\in
\mathcal{R}^V_{KEY}$, we denote by $view^{P(SK_P),
V(SK_V)}_{\mathcal{A}}(1^n, X, z, PK_P, PK_V)$ the random variable
describing the view of $\mathcal{A}$ specific to $(PK_P, PK_V)$,
which includes its random tape, the auxiliary string $z$, the
(specific) $(PK_P, PK_V)$, and all messages it receives from the
instances of $P(1^n, SK_P)$ and $V(1^n, SK_V)$ in the
proof stages. 

\end{description}

Note that in all cases, the honest prover and verifier instances
answer messages from $\mathcal{A}$ promptly. We stress that in
different left or right sessions the honest prover and verifier
instances use independent random-tapes in the proof stages.
\emph{The adversary's goal is to complete a right session with
statement  different from that of any left session, for which the
verifier accepts even if the adversary actually does not know a
witness for the statement being proved.} 



\section{Formulating CNMZK in the Public-Key Model,
Revisited}\label{CNMrevisited}

Traditional CNMZK formulation  roughly is the following: for any
PPT CMIM adversary $\mathcal{A}$ of traditional input selecting
capability (as clarified
in Section \ref{CMIMBPK}), 
 there exists a PPT
simulator/extractor $S$ such that $S$ outputs the following: (1) A
simulated transcript that is indistinguishable from the real view
of the CMIM adversary in its CMIM attacks. (2) For a successful
right session on a common input $\hat{x}$  different from those of
left sessions, $S$ can output a corresponding
$\mathcal{NP}$-witness of $\hat{x}$.

The requirement (1) intuitively captures that any advantage of
$\mathcal{A}$ can get from concurrent left and right interactions
can also be got by $S$ itself alone without any interactions, i.e,
$\mathcal{A}$ gets no extra advantage by the CMIM attacks. The
requirement (2) intuitively captures that for any different
statement that $\mathcal{A}$ convinces of $V$ in one  of right
sessions, $\mathcal{A}$ must ``know" a witness.

 The formulations of CNM in
the public-key model in existing works
(\cite{Z03,OPV06,DDL06,OPV07}) essentially  directly bring the
above traditional  CNM formulation  into the public-key setting,
but with the following difference: $S$ will simulate the
key-generation phases of all honest verifiers. Put in other words,
in its simulation/extration $S$ actually takes the corresponding
secret-keys of honest verifiers.


 We start clarifying  the subtleties of CNM in the public-key model
by showing a CMIM attack on the CNMZK in the BPK model proposed in
\cite{DDL06}. The CMIM attack allows the CMIM adversary to
successfully convince the honest verifier of some $\mathcal{NP}$
statements but without knowing any witness to the
statement being proved.            

\subsection{CMIM  attacks on  the CNMZK proposed in  \cite{DDL06}}
\label{attack}

Let us first recall the protocol structure of the  protocol of
\cite{DDL06}.

\begin{description}
\item [Key-generation.] Let $(KG_0, Sig_0, Ver_0)$ and $(KG_1,
Sig_1, Ver_1)$ be two signature schemes that secure against
adaptive chosen message attacks.   On a security parameter $1^n$,
each verifier $V$
 randomly generates two pair $(verk_0, sigk_0)$ and $(verk_1, sigk_1)$ by running $KG_0$ and $KG_1$ respectively, where $verk$ is the signature verification key
 and $sigk$ is the signing key.   $V$ publishes
$(verk_0, verk_1)$
 as its public-key while keeping $sigk_b$ as its secret-key for a randomly chosen $b$  from $\{0,
 1\}$ ($V$ discards $sigk_{1-b}$). The prover does not possess
 public-key.

\item [Common input.] An element $x\in \mathcal{L}$ of length
$poly(n)$, where $\mathcal{L}$ is an $\mathcal{NP}$-language that
admits $\Sigma$-protocols.

\item[The main-body of the protocol.] The main-body of the
protocol consists of the following three phases:
\begin{description}

\item[Phase-1.] The verifier $V$ proves to $P$ that it knows
either $sigk_0$ or   $sigk_1$, by executing the (partial
witness-independent) $\Sigma_{OR}$-protocol \cite{CDS94} on
$(verk_0, verk_1)$
 in which $V$ plays the role of knowledge prover.
    Denote by $a_{V}$, $e_{V}$, $z_{V}$, the first-round, the second-round
and the third-round message of the $\Sigma_{OR}$-protocol of this
phase respectively. Here $e_{V}$ is the random challenge sent by
the prover to the verifier.

 If $V$ successfully finishes the
 $\Sigma_{OR}$-protocol of this phase and $P$ accepts, then goto Phase-2.
 Otherwise, $P$ aborts.

 \item[Phase-2.] $P$ generates a key pair $(sk, vk)$ for a one-time
 strong signature scheme. Let $COM$ be a  commitment scheme. The prover randomly selects
 random  strings $s, r\in \{0, 1\}^{poly(n)}$, and computes $C=COM(s, r)$
 (that is, $P$ commits to $s$ using randomness $r$). Finally, $P$ sends $(C, vk)$  to
 the verifier $V$.

\item [Phase-3.] By running a $\Sigma_{OR}$-protocol, $P$ proves
to $V$ that it knows either a witness $w$ for $x\in \mathcal{L}$
OR the  value committed in $C$ is a signature on the message of
$vk$ under either $verk_0$ or $verk_1$.  Denote by $a_P, e_P,
z_P$, the first-round, the second-round and the third-round
message of the $\Sigma_{OR}$ of Phase-3. Finally, $P$ computes a
one-time strong signature $\delta$ on the whole transcript with
the signing key $sk$ generated in Phase-2.

\item[Verifier's decision.] $V$ accepts if and only if the
$\Sigma_{OR}$-protocol of Phase-3 is accepting, and  $\delta$ is a
valid signature on the whole transcript under $vk$.

\end{description}
\end{description}

\textbf{Note:} The actual implementation of the DDL protocol
combines rounds of the above protocol. But, it is easy to see that
round-combination does not invalidate the following attacks.

\subsubsection{The CMIM  attack}

We  show a special CMIM attack in which the adversary
$\mathcal{A}$ only participate the right concurrent interactions
with honest verifiers (i.e., there are no concurrent left
interactions in which $\mathcal{A}$ concurrently interacts with
honest provers).

The following CMIM  attack  enables $\mathcal{A}$ to malleate the
interactions of Phase-1 of one session into a successful
conversation of another concurrent session for different (but
verifier's public-key related) statements without knowing any
corresponding $\mathcal{NP}$-witnesses. 

Let $\hat{L}$ be any $\mathcal{NP}$-language admitting a
$\Sigma$-protocol that is denoted by $\Sigma_{\hat{L}}$ (\emph{in
particular, $\hat{L}$ can be an empty set}). For an honest
verifier $V$ with its public-key $PK=(verk_0, verk_1)$, we define
a new language $\mathcal{L}=\{(\hat{x}, verk_0, verk_1)| \exists w
\ s.t. \ (\hat{x}, w)\in \mathcal{R}_{\hat{L}} \ \text{OR}\
w=sigk_b  \ \text{for} \ b \in \{0, 1\}\}$. Note that for any
string $\hat{x}$ (whether $\hat{x}\in \hat{L}$ or not), the
statement ``$(\hat{x}, verk_0, verk_1) \in \mathcal{L}$'' is
always true as $PK=(verk_0, verk_1)$ is honestly generated. Also
note that $\mathcal{L}$ is a language that admits
$\Sigma$-protocols (as $\Sigma_{OR}$-protocol is itself a
$\Sigma$-protocol). Now, we describe the concurrent interleaving
and malleating attack,  in which $\mathcal{A}$ successfully
convinces the honest verifier of the statement ``$(\hat{x},
verk_0, verk_1) \in \mathcal{L}$'' for \emph{any arbitrary}
$poly(n)$-bit string $\hat{x}$ (\emph{even when $\hat{x}\not \in
\hat{L}$}) by concurrently interacting with $V$ (with public-key
$(verk_0, verk_1)$) in two sessions as follows.

\begin{enumerate}
\item $\mathcal{A}$ initiates the first session with $V$.  After
receiving the first-round message, denoted by  $a^{\prime}_V$, of
the $\Sigma_{OR}$-protocol  of Phase-1 of the first session on
common input $(verk_0, verk_1)$ (i.e.,  $V$'s public-key),
$\mathcal{A}$ suspends the first session.

\item  $\mathcal{A}$ initiates a second session   with $V$,  and
works just as the honest prover does in Phase-1 and Phase-2 of the
second session. We denote by $C, vk$ the Phase-2 message of the
second session, where $C$ is the commitment  to a random string
and $vk$ is the verification key of the one-time strong signature
scheme generated by $\mathcal{A}$ (\emph{note that $\mathcal{A}$
knows the corresponding signing key $sk$ as $(vk, sk)$ is
generated by itself}). When $\mathcal{A}$ moves into Phase-3 of
the second session and needs to send $V$ the first-round message,
denoted by $a_P$, of the $\Sigma_{OR}$-protocol of Phase-3 of the
second session \emph{on common input $(\hat{x}, verk_0, verk_1)$},
$\mathcal{A}$ does the following:

\begin{itemize}

\item $\mathcal{A}$ first runs the  SHVZK simulator of
$\Sigma_{\hat{L}}$ (i.e.,  the $\Sigma$-protocol for $\hat{L}$)
\cite{D03}  on $\hat{x}$ to get a simulated conversation, denoted
by $(a_{\hat{x}}, e_{\hat{x}}, z_{\hat{x}})$, for the
(\emph{possibly false})  statement ``$\hat{x} \in \hat{L}$".

\item $\mathcal{A}$ runs the SHVZK simulator of the
$\Sigma$-protocol for showing that the value committed in   $C$ is
a signature on $vk$ under one of $(verk_0, verk_1)$ to get a
simulated conversation, denoted by $(a_C, e_C, z_C)$.

\item $\mathcal{A}$ sets $a_P=(a_{\hat{x}}, a^{\prime}_V, a_C)$
and sends $a_P$ to $V$ as the first-round message of the
$\Sigma_{OR}$-protocol of Phase-3 of the second session, where
$a^{\prime}_V$ is the one received by $\mathcal{A}$ in the first
session.

\item After receiving the second-round message of Phase-3 of the
second session, i.e., the random challenge $e_P$ from $V$,
$\mathcal{A}$ suspends the second session.
\end{itemize}

\item $\mathcal{A}$ continues the first session, and sends
$e^{\prime}_V=e_P\oplus e_{\hat{x}} \oplus e_C$ as the
second-round message of the $\Sigma_{OR}$-protocol of Phase-1 of
the first session.

\item After receiving the third-round message of the
$\Sigma_{OR}$-protocol of Phase-1 of the first session, denoted by
$z^{\prime}_V$, $\mathcal{A}$ suspends the first session again.

\item $\mathcal{A}$ continues the execution of the  second session
again, sends to  $z_P=((e_{\hat{x}}, z_{\hat{x}}), (e^{\prime}_V,
z^{\prime}_V), (e_C, z_C))$ to $V$  as the third-round message of
the $\Sigma_{OR}$-protocol of  the second session.

\item Finally, $\mathcal{A}$ applies $sk$ on the whole transcript
\emph{of the second session} to get a (one-time strong) signature
$\delta$, and sends $\delta$ to $V$

\end{enumerate}

 Note that  $(a_{\hat{x}}, e_{\hat{x}},
z_{\hat{x}})$ is an accepting conversation for the (possibly
false) statement ``$\hat{x}\in \hat{L}$",  $(a^{\prime}_V,
e^{\prime}_V, z^{\prime}_V)$ is an accepting conversation for
showing the knowledge of  either $sigk_0$ or $sigk_1$, $(a_C, e_C,
z_C)$ is an accepting conversation for showing that the value
committed in $C$ is a signature on $vk$ under one of $(verk_0,
verk_1)$. Furthermore, $e_{\hat{x}}\oplus e^{\prime}_V\oplus
e_C=e_P$, and $\delta$ is a valid (one-time strong) signature on
the transcript of the second session.This means that, from the
viewpoint of $V$, $\mathcal{A}$ successfully convinced $V$ of the
statement ``$(\hat{x}, verk_0, verk_1) \in \mathcal{L}$''
 in the second session
  \emph{but without knowing any
corresponding $\mathcal{NP}$-witness}!



\subsection{Reformulating  CNMZK in the BPK
model}\label{CNMZKdef}


In light of the above CMIM attacks, we highlight a key difference
between the CMIM setting in the public-key model and the CMIM
setting in the standard model.

\textsc{The key difference:} For CMIM setting in the standard
model, honest verifiers are PPT algorithms. In this case, normal
CNM formulation only considers the extra advantages the CMIM
adversary can get from concurrent left sessions, as the actions of
honest verifiers in right sessions can be efficiently emulated
perfectly; But, for CMIM setting in the public-key model, the
honest verifier possesses secret value (i.e, its secret-key) that
can \emph{NOT} be computed out  efficiently from the public-key.
In other words, in this case an CMIM adversary can get extra
advantages both from the left sessions and \emph{from the right
sessions}. This is a crucial difference between CMIM settings for
standard model and public-key model, which normal formulation of
CNM does not capture. The CMIM attack on the protocol of
\cite{DDL06} clearly demonstrates this difference.


With the above key difference in mind, we investigate
reformulating the CNM notion in the public-key model. Above all,
besides requiring the ability of simulation/extraction, we need to
  mandate that for any CMIM-adversary the witnesses extracted for right sessions are  ``\emph{independent}"
of the  secret-key used by the simulator/extractor $S$ (who
emulates
 honest verifiers in the simulation/extraction). Such property is named \textsf{concurrent non-malleable
 knowledge-extraction independence (CNMKEI)}.  CNMKEI  is formulated by extending the formulation of concurrent
 knowledge-extraction (CKE) of \cite{YYZ07} into the more complicated CMIM setting
 (the CKE notion is formulated with adversaries only
 interacting with honest verifiers but without interacting with provers). 
Roughly, the CNMKEI is formulated as follows.

\textsc{CNMKEI in the public-key model:} We require that for any
PPT CMIM-adversary $\mathcal{A}$ in the BPK model, there exists a
PPT simulator/extractor $S$ such that the following holds:
$\Pr[\mathcal{R}(\widehat{W}, SK_V, str)] =1$ is negligibly close to
$\Pr[\mathcal{R}(\widehat{W}, SK^{\prime}_V, str)]=1$ for any polynomial-time
computable relation $\mathcal{R}$, where $SK^{\prime}_V$ is some element
\emph{randomly and independently} distributed over the space of
$SK_V$, $str$ is the simulated transcript indistinguishable from the
real view of $\mathcal{A}$, and $\widehat{W}$ are the joint
witnesses extracted to successful right sessions in $str$. Here,
for some right session that is aborted (due to CMIM adversary
abortion or verifier verification failure) or is of common input
identical to that of one left session, the corresponding witness
to that right session is set to be a special symbol $\perp$.

The formal formulation of the reformulated CNMZK definition in the
BPK model is presented below: 

\begin{definition}  [CNMZK in the public-key
model]\label{cke} 
  We  say that  a protocol $\langle P, V\rangle$ is
  \textsf{\textup{concurrently non-malleable zero-knowledge}}
  in the BPK  model w.r.t. a class of admissible languages $\mathcal{L}$ 
   and some
  key-validating relations  $\mathcal{R}^P_{KEY}$ and $\mathcal{R}^V_{KEY}$,
if for  any positive polynomial $s(\cdot)$, any $s$-CMIM adversary
$\mathcal{A}$ defined in Section \ref{CMIMBPK}, there exist a pair
of (expected) polynomial-time algorithms $S=(S_{KEY}, S_{PROOF})$
(the simulator) and $E$ (the extractor) such that for any
sufficiently large $n$, any auxiliary input $z \in \{0, 1\}^*$, any $\mathcal{NP}$-relation $\mathcal{R}_L$ (indicating an admissible language $L\in \mathcal{L}$),
and any polynomial-time computable relation $\mathcal{R}$ (with
components drawn from $\{0, 1\}^* \cup \{\perp\}$), the following
hold, in accordance with the experiment
\textbf{$\textsf{Expt}_{\textup{CNM}}(1^n, X, z)$} described below
(page \pageref{experiment}):


 \begin{table}[!h]
\begin{center}

\begin{tabular} {|c|}
 \hline \label{experiment}

 \textbf{$\textsf{Expt}_{\text{CNM}}(1^n, X, z)$}\\ 

\begin{minipage}[t] {5.7in} 
\vspace{0.1cm}

\textbf{Honest prover key-generation:}

 $(PK_P, SK_P)\longleftarrow
P_1(1^n)$. Denote by $\mathcal{K}_L$ the set of all legitimate
public-keys generated by $P_1(1^n)$. \emph{Note that the execution
of $P_1$ is independent from the simulation below. In particular,
only the public-key $PK_P$ is passed on to the simulator. }
\\

\textbf{The simulator $S=(S_{KEY}, S_{PROOF})$:}

$(PK_V, SK_V, SK^{\prime}_V)\longleftarrow S_{KEY}(1^n)$, where
the distribution of $(PK_V, SK_V)$  is identical with that of the
output of the key-generation stage of the honest  verifier $V_1$,
$\mathcal{R}^V_{KEY}(PK_V, SK_V)=\mathcal{R}^V_{KEY}(PK_V,
SK^{\prime}_V)=1$ and the distributions of $SK_V$ and
$SK^{\prime}_V$ are identical and \emph{independent}. In other
words, $SK_V$ and $SK^{\prime}_V$ are two random and independent
secret-keys corresponding to $PK_V$.
 \\


 $(str, sta) \longleftarrow S_{PROOF}^{\mathcal{A}(1^n,\ X, \ PK_P,\ PK_V, \  z)}(1^n, X,  PK_P,  PK_V,
   SK_V, z)$. That is,   on inputs $(1^n, X, PK_P, PK_V, SK_V, z)$ and with
oracle access to $\mathcal{A}(1^n, X,  PK_P, PK_V, z)$ (defined in
accordance with the   experiment
$\textsf{Expt}^{\mathcal{A}}_{CMIM}(1^n, X, W, z)$ described  in
Section  \ref{CMIMBPK}), the simulator $S$ outputs a simulated
transcript $str$,   and some state information $sta$  to be
transformed to the knowledge-extractor $E$. \emph{Note that $S$
does not know the secret-key $SK_P$ of honest prover, that is, $S$
can emulate the honest prover only from its public-key $PK_P$. }
\\

For any $X\in L^{s(n)}$ and $z\in \{0, 1\}^*$, we denote by
$S_1(1^n, X, z)$ the random variable  $str$ (in accordance with
above processes of $P_1$,  $S_{KEY}$ and $S_{PROOF}$). For any
$X\in L^{s(n)}$, $PK_P \in \mathcal{K}_P$  and $(PK_V, SK_V)\in
\mathcal{R}^V_{KEY}$ and any $z\in \{0, 1\}^*$, we denote by
$S_1(1^n, X, PK_P, PK_V,  SK_V, z)$ the random variable describing
the first output of $S_{PROOF}^{\mathcal{A}(1^n,\ X, \ PK_P, \
PK_V, \ z)}(1^n, X, PK_P, PK_V,  SK_V, z)$ (i.e., $str$ specific
to $(PK_P, PK_V, SK_V)$).
\\

\textbf{The knowledge-extractor $E$:}

$\widehat{W} \longleftarrow E(1^n, sta, str)$. On $(sta, str)$,
$E$ outputs a list of witnesses to (different right) statements
whose validations are successfully conveyed in right sessions in
$str$, where each of these statements  is different from the
statements of
left sessions. 

\end{minipage}
\\
\hline
\end{tabular}
\end{center}

\end{table}

\begin{itemize}
\item \textbf{\textup{ Simulatability.}}
The following ensembles are  indistinguishable: \\
 $\{S_1(1^n, X, PK_P,  PK_V, SK_V, z)\}_{X\in L^{s(n)}, PK_P\in \mathcal{K}_P,
 (PK_V, SK_V)\in \mathcal{R}^V_{KEY}, z\in\{0, 1\}^*}$
 and \\
   $\{view^{P(SK_P), V(SK_V)}_{\mathcal{A}}(1^n, X, PK_P, PK_V,  z)\}_{X\in L^{s(n)},
   PK_P\in \mathcal{K}_P, (PK, SK)\in R_{KEY}, z\in\{0,
  1\}^*}$ (defined in accordance with the   experiment $\textsf{Expt}^{\mathcal{A}}_{CMIM}(1^n, X, W,
z)$ described  in Section \ref{CMIMBPK}). This in particular
implies
 the probability
ensembles $\{S_1(1^n, X,  z)\}_{X\in \\ L^{s(n)}, z\in\{0, 1\}^*}$
and $\{view_{\mathcal{A}}(1^n, X, z)\}_{X\in L^{s(n)}, z\in\{0,
1\}^*}$ are indistinguishable.


%



\item \textbf{\textup{Secret-key independent
knowledge-extraction.}}  $E$, on inputs  $(1^n, str, sta)$,
 outputs witnesses to all (different  right) statements successfully proved in
accepting right sessions in $str$ (with each of the statements
different from those of left sessions). Specifically, $E$ outputs
a list of strings $\widehat{W}=(\hat{w}_1, \hat{w}_2, \cdots,
\hat{w}_{s(n)})$, satisfying the following:

\begin{itemize}
\item $\hat{w}_i$ is set to be $\perp$, if the $i$-th  right
session in $str$ is not accepting (due to abortion or verifier
verification failure) or the common input of the $i$-th right
session is identical with that of one of left sessions, where
$1\leq i \leq s(n)$.

\item \textsf{\textup{Correct knowledge-extraction for
(individual) statements:}} In any other cases, with overwhelming
probability $(\hat{x}_i,
\hat{w}_i)\in \mathcal{R}_\mathcal{L}$,  where 
 $\hat{x}_i$ is the statement  selected by $P^*$ for the $i$-th
 right
session in $str$  and $\mathcal{R}_{\mathcal{L}}$ is the
$\mathcal{NP}$-relation for the admissible language  $L\in \mathcal{L}$ set by $P^*$ for right sessions  in
$str$.

\item \textsf{\textup{concurrent non-malleable knowledge
extraction independence (CNMKEI):}} \
$\Pr[\mathcal{R}(SK_V, \widehat{W}, str)=1]$ is negligibly close
to $\Pr[\mathcal{R}(SK^{\prime}_V, \widehat{W}, str)=1]$. This in
particular implies that the distributions of $(PK_V, SK_V, str)$
and $(PK_V, SK^{\prime}_V, str)$ are indistinguishable (by
considering $PK_V$ encoded in $\widehat{W}$).


\end{itemize}
The probabilities are taken over the randomness of $P_1$, the
randomness of $S$ in the key-generation stage (i.e., the
randomness for generating $(PK_V, SK_V, SK^{\prime}_V)$) and in
all proof stages, the randomness of $E$, and
the randomness of $\mathcal{A}$. 
\end{itemize}

\end{definition}

Note that the above CNM formulation in the public-key model
 implies both concurrent ZK for concurrent prover security in the public-key model
 (note that $S$ emulates
 the honest prover without knowing its secret-key),
  and  concurrent knowledge-extraction for concurrent verifier security
   in the public-key model formulated in
  \cite{YYZ07}. The  CNM formulation
  follows the simulation-extraction approach of
  \cite{PR05}, and extends the CKE formulation of \cite{YYZ07} into the more complex CMIM setting.
  We remark that, as clarified,  mandating the CNMKEI property
  is crucial for correctly formulating CNM security in the
  public-key model. We also note that the above CNMZK definition in the BPK model  can be
  trivially extended to a tag-based formalization version

\subsection{Discussions and
clarifications}\label{CNMZKclarification}


\textbf{Existing CNM formulations in the public-key model do not
capture CNMKEI.}
 The CNM formulation in the work \cite{OPV06}
uses  the   indistinguishability-based approach of \cite{PR05}.
Specifically, in the CNM formulation of \cite{OPV06},  two
experiments are defined (page 19 of \cite{OPV06}): a real
experiment w.r.t. a real public-key of an honest verifier (here,
denoted $PK_V$), in which a CMIM adversary mounts CMIM attacks; a
simulated experiment run by a simulator/extractor $S$ w.r.t. a
simulated public-key (here, denoted $PK_S$), in which $S$ accesses
$\mathcal{A}$ and takes a simulated secret-key $SK_S$. The CNM is
then formulated as follows: the distribution of all witnesses used
by $\mathcal{A}$ in right sessions in the real experiment is
indistinguishable from the distribution of the witnesses used by
$\mathcal{A}$ in right sessions in the simulated experiment.
\emph{Note that \cite{OPV06} does not require the
simulator/extractor to output a simulated indistinguishable
transcript.} That is, the    CNM formulation of \cite{OPV06} does
\emph{not} automatically imply concurrent zero-knowledge.

It appears that the CNM formulation of \cite{OPV06} has already
dealt with the issue of knowledge-extraction independence.  But, a
careful investigation shows that it does not.  The reason is as
follows:

Firstly, in the real experiment the statements selected by the
CMIM adversary $\mathcal{A}$ for both left and right sessions can
be maliciously related to $PK_V$ (e.g., some function of $PK_V$),
and thus the witnesses extracted for right sessions of the real
experiment could be potentially dependent on the secret-key $SK_V$
used by honest players. Note that, as witnessed by the above
concurrent interleaving and malleating attack 
on the CNMZK protocol of \cite{DDL06}, when extracted witnesses
are maliciously dependent on $SK_V$ knowledge-extraction does not
necessarily capture the intuition that $\mathcal{A}$ does ``know" the
witnesses extracted. Similarly, as in the simulated experiment $S$
uses $SK_S$ in simulation/extraction, the witness extracted in the
simulated experiment could also be maliciously dependent on
$SK_S$. That is, both the witnesses extracted in real experiment
and in the simulated experiment may be maliciously dependent on
$SK_V$ and $SK_S$ respectively, \emph{but the distributions of
them still can be indistinguishable as the distributions of $SK_V$
and $SK_S$ are identical!}\footnote{We note that the CNMZK definition in \cite{OPV06} was modified in 
\cite{OPV07} in March 2007, after we revealed this observation in \cite{YYZ07} in January of 2007 (the preliminary version of \cite{YYZ07} was submitted to CRYPTO 2007).}

The CNMZK formulations in the subsequent works of
\cite{DDL06,OPV07} are essentially the traditional CNMZK
formulation following the simulation/extraction approach, which is
incomplete for correctly capture CNM security in the public-key
model as clarified above.

\textbf{CNM with full adaptive input selection.} The above CNMZK
formulation does not explicitly specify the input-selecting
capabilities of  the CMIM adversary. According to the
clarifications presented  in Section \ref{fulladaptive}, there are
four kinds of CNM security to be considered: CNM security against
\textsf{CMIM with predetermined inputs}, CNM security against
\textsf{CMIM with adaptive input selection}, CNM security against
\textsf{CMIM with predetermined left-session inputs but full
adaptive input selection on the right}, and CNM security against
\textsf{CMIM with full adaptive input selection}.

We briefly note that no previous protocols in the BPK model were
proved to be CNM-secure against even \textsf{CMIM with
predetermined left-session inputs but full adaptive input
selection on the right} (i.e., the inputs to left sessions are
predetermined and the CMIM adversary only sets inputs to right
session in the fully adaptive way),   needless to say to be CNM
secure against \textsf{CMIM with full adaptive input selection}.
Specifically, the standard simulation-extraction paradigm for
showing   CNM security fails, in general, when the CMIM adversary
is allowed the capability of full adaptive input selection.

In more detail, the standard simulation-extraction paradigm  for
establishing  CNM security works as follows: the simulator first
outputs an indistinguishable
simulated  transcript; and then extracts 
 the witnesses to (different)  inputs of  successful  right
sessions appearing in the simulated transcript, \emph{ one by one
sequentially},  by applying some assured underlying
knowledge-extractor. This paradigm can work for CMIM adversary
with the capability of traditional adaptive input selection,  as
the input to each right session is fixed at the beginning of the
right session; Thus, applying knowledge-extractor on the right
session does not change the statement of the session, which has
appeared and is fixed
  in the simulated transcript. 

 But, for CMIM adversary
of fully adaptive input selection, the standard
simulation-extraction paradigm fails in general in this case.  In
particular, considering the adversary always sets inputs to  right
sessions only at the last message of each right session, such case
applies to both of the two illustrative natural protocol examples
presented in Section \ref{fulladaptive}: composing coin-tossing
and NIZK, and the Feige-Shamir-ZK-like protocols. In this case,
when we apply knowledge-extractor on a successful right session,
the statement of this session will however also be changed,  which
means that the extractor may never extract witness to the same
statement appearing and being fixed in the simulated transcript.
More detailed clarifications are given in Section \ref{DEFCNMCT},
following the definition of concurrent non-malleable coin-tossing
in the BPK model.

\textbf{On the possibility of CNMZK with adaptive input selection
in the BPK model.} The possibility of CNMZK with adaptive (not
necessarily to be fully adaptive) input selection in the BPK model
turns also out to be a quite subtle issue. In particular, we note
that (traditional) adaptive input selection was highlighted for
the CNMZK in \cite{OPV06}, but the updated version of \cite{OPV07}
are w.r.t. predetermined prover inputs (such subtleties were not
clarified in \cite{OPV06,OPV07}. It appears that, as noted
recently in \cite{P06}, the existence of CNMZK with adaptive
(needless to say fully adaptive)  input selection in the BPK model
might potentially violate Lindell's impossibility results on
concurrent composition with adaptive input selection
\cite{Ljoc,L03focs}. This raised the question that: whether
constant-round CNMZK protocols (particularly in accordance with
our CNMZK formulation) with adaptive input selection exists in
the BPK model (or, whether it is possible at  least)?  


A careful investigation shows that constant-round CNMZK with
adaptive input selection could still be possible in the BPK model,
and actually our work does imply such protocols with the strongest
\emph{full} adaptive input selection. Below, we give detailed
clarifications in view of Lindell's impossibility results
of \cite{Ljoc,L03focs}. 
Lindell's impossibility results of \cite{Ljoc,L03focs} hold  for
concurrent (self or general) composition of protocols securely
realizing (large classes of)  functionalities enabling (bilateral)
bit transmission. The Zero-Knowledge functionality  $((x, w),
\lambda)\mapsto (\lambda, (x, R(x, w)))$ enables unilateral bit
transformation from prover to verifier. But, when a CNMZK protocol
\emph{in the plain model} is considered, where the CMIM adversary
can play both the role of prover and the role of the verifier
(note that the honest verifier can be perfectly emulated by the
CMIM adversary in the plain model), it actually amounts to realize
an extended version of
 ZK functionality \emph{with interchangeable roles} that does enable
 bilateral bit transformation in this case. This implies that CNMZK
 with adaptive input selection is impossible in the plain model.

 The ZK (not necessarily CNMZK) protocol for an $\mathcal{NP}$-language $\mathcal{L}$  in the BPK model essentially
 amounts to securely realizing the following functionality: $((x, w), (PK_V, SK_V))
 \mapsto ((PK_V, \mathcal{R}^V_{KEY}( PK_V,\\  SK_V)),
(x, \mathcal{R}_L(x, w)))$ that enables bilateral bit
transmission. This means that when adaptive input selection is
allowed both for prover inputs and \emph{verifier's keys}, which
implies the verifier's keys and thus the public file output by the
key-generation stage are not fixed but are set accordingly by the
CMIM adversary in order to transmit bits from honest verifiers to
honest provers, even concurrent ZK (needless to
say CNMZK) may  not exist in the BPK model!  
 We highlight some key points that still could allow  the possibilities of  CNMZK with adaptive input selection
   in the BPK model:

   \begin{itemize}
   \item \textbf{Disabling bit transformation from honest verifiers to other players:}
   Note that: in key-generation stage, the keys of \emph{honest}
verifiers are generated independently by the honest verifiers
themselves and cannot be set adaptively by the CMIM adversary; In
the proof stages, the keys of honest verifiers (actually all keys
in the public file)  cannot be  modified by the CMIM adversary, as
we assume the public file used in the proof stages remains the
same output at the end of key-generation stage;  Furthermore, in
the BPK setting we assume the role of honest verifiers with
honestly generated keys is fixed. That is,  honest verifiers may
prove the knowledge of their corresponding secret-keys, but they
never prove anything else.

 Putting all together, it means that honest verifiers
instantiated with their public-keys cannot be  impersonated and
emulated by the CMIM adversary, and their inputs (i.e., the keys
generated in key-generation stage and then fixed and remaining
unchanged for proof stages) and their prescribed  actions and
player role in the proof stages are not influenced by the CMIM
adversary. This disables bit transmission from honest verifiers to
other players, which implies that the existence of CNMZK with
adaptive input selection in the BPK model could  still not violate
Lindell's impossibility results.

\item  \textbf{Disabling bit transformation from other players to
honest provers:} For a protocol in the BPK model, the public-keys
registered by honest provers and the public-keys registered by
honest verifiers can be of different types, and the use of
honest-prover keys and the use of honest-verifier keys in protocol
implementation can also be totally different. Such differences can
be on the purpose of protocol design,  as demonstrated  with our
CNMCT implementation.
 Then, for
honest provers of fixed role in the BPK model, though the CMIM
adversary can enable, by adaptive input selection, bit
transmissions from honest provers to other players, but, in the
BPK model, the CMIM adversary may not enable bit transmissions
from other players to honest provers.



\item 
\textbf{Concurrent self composition vs. concurrent general
composition in the BPK model:}
We further consider a more general case for any two-party protocol
$\langle P, V\rangle$ in the BPK model.  Suppose there are some
players of fixed role, and some players of interchangeable roles
(i.e., players who can serve both as prover and as verifier). The
direct way for a player in the BPK model to be of interchangeable
roles is to register a pair of keys $(PK_P, PK_V)$ and to
explicitly indicate its role, i.e., prover or verifier,  in the
run of each session. Then, according to the analysis of
\cite{Ljoc,L03focs}, the run of any arbitrary external protocol
executed among players of interchangeable roles can be emulated,
by a CMIM adversary capable of adaptive input selection, in the
setting of concurrent self composition of the protocol $\langle P,
V\rangle$ among those players. But, the external protocol
executions involving honest players of fixed roles, however, are
not necessarily be able to be emulated by self-composition of the
protocol involving  the honest players of fixed roles. This
implies that, as long as there are honest players of fixed roles
in the BPK model, concurrent self-composition with adaptive input
selection in the BPK system  does not necessarily imply concurrent
general composability.

\end{itemize}

\textsf{A tradeoff.} The above clarifications also pose a tradeoff
between players' roles and their CNM security levels in the BPK
model: For stronger CNM security  of adaptive input selection,
honest players in the BPK model need to be of fixed roles; Of
course, honest players can also choose to be of interchangeable
roles for their own convenience, but with the caveat  that CNM
security against CMIM of adaptive input selection may lose (though
CNM with predetermined inputs can still remain). In other words,
whether to be of fixed role or interchangeable role  can be at the
discretion of each honest player in the BPK model. If one is
interested with the  stronger CNM security against CMIM of (full)
adaptive input selection, it is necessary for it to be of fixed
role. A typical scenario of this case is: this player is a server,
who normally plays the same role and takes higher priority of
stronger security over Internet; However, if one is interested in
the convenience of interchangeable role, it can simply register a
pair of keys $(PK_P, PK_V)$ and explicitly indicate its role in
the run of  each session, but with the caveat that its CNM
security against CMIM of adaptive input selection may lose. 

\section{Constant-Round CNM Coin-Tossing  in the  BPK
Model}\label{CNMCT}

Coin-tossing is one of the first and more fundamental protocol
problem in the literature \cite{C01}. In its simplest form, the
task calls for two mutually distrustful parties to generate a
common random string \cite{B82}. In this section, we formulate and
achieve constant-round concurrent non-malleable coin-tossing in
the more complex CMIM setting in the BPK model, which can be used
to move concurrent  non-malleable cryptography from common random
string model into the weaker
BPK model. 

\subsection{Definition of CNM  coin-tossing in the BPK
model}\label{DEFCNMCT} Let $\langle L, R\rangle$ be a coin-tossing
protocol between a left-player $L$ and a right-player $R$. (We abuse
the notations $L$ and $R$ in this  section. Specifically, $L$
stands for the left-player and in some context we may  explicitly
indicates $L$ to be a language, $R$ stands for the right-player
and in some context we may  explicitly indicates $R$ to be a
relation.) The CMIM setting for coin-tossing in the BPK model can
be slightly adapted (actually simplified) from the CMIM setting
for CNMZK in the BPK
model (formulated in Section \ref{CMIMBPK}). 
  Note that coin-tossing amounts to the
functionality: $(\lambda, \lambda) \mapsto (r, r)$, where $r$ is a
random string. As players possess no inputs in the coin-tossing
functionality,  the issue of adaptive input selection does not
apply to coin-tossing. But, as we shall see, the CNMCT formulated and achieved herein can be used to transform  CNM cryptography from CRS model to the BPK model with fully adaptive input selection.

To formulate CNMCT in the
complex CMIM setting, the rough idea is: 
  for any CMIM adversary $\mathcal{A}$ there exists a PPT simulator $S$ such
that: (1) $S$ outputs a simulated transcript $str$
indistinguishable from the real view of $\mathcal{A}$, together
with some state information $sta$; (2) $S$ can set, at its wish,
``\emph{random} coin-tossing outputs'' for all (left and right)
sessions in $str$, in the sense that  $S$ learns the corresponding
trapdoor information (included in 
 $sta$) of
the coin-tossing
 output of each session.
  Intuitively, such formulation implies  the
 traditional simulation-extraction CNM security.
 But, with the goal of transforming CNM
 cryptography from CRS model into the weaker BPK model in mind, some
 terms need to be further
  deliberated.

  Above all, we need
  require 
  the combination of 
  $str$ and  
   $sta$ should be  independent of the secret-key emulated and used by the simulator.
   This is necessary to guarantee that  $\mathcal{A}$ knows what it claims to know
   in its CMIM attack.

   Secondly, we should mandate the ability of
  \emph{online} setting coin-tossing outputs of all sessions appearing in $str$, in
  the sense that $S$ sets the coin-tossing outputs and  the corresponding  trapdoor
  information (encoded in $sta$)
   in an online way  at the same
  time of forming the  $str$. This is critical to
  guarantee CNM security against CMIM with full adaptive input
  selection.

  Finally, we need to make clear the meaning of
  ``random coin-tossing outputs''. One formulation 
   is to require
  that all coin-tossing outputs are independent random strings. Such formalization rules
out the natural copying strategy by definition, and thus is too
strong to capture naturally secure 
 protocols. On the other hand, in order to  allow the copying strategy to the CMIM, 
 an alternative relaxed formulation  is to only require
that the coin-tossing output of each \emph{individual} 
session is random. But, this alternative formalization is too week
to rule out naturally insecure protocols  (for instance, consider
that the CMIM manages to set the outputs of some
 sessions to be maliciously correlated and even to be  identical). 
The right formulation should essentially be: \emph{the
 coin-tossing
 output of each left (resp., right) session  is either
independent of the outputs of all other sessions OR  copied from
the output of one right (resp., left) session on the opposite CMIM
 part; furthermore, the output of each session in one
CMIM part can be copied into the opposite  CMIM part at most
once.}

\textbf{Legitimate CRS-simulating algorithm $\mathcal{M}_{CRS}$.}
Let $(r, \tau_r)\longleftarrow \mathcal{M}_{CRS}(1^n)$, where
$\mathcal{M}_{CRS}$ is a PPT algorithm. The PPT algorithm
$\mathcal{M}_{CRS}$ is called a legitimate CRS-simulating
algorithm with respect to a polynomial-time computable
CRS-trapdoorness validating relation $\mathcal{R}_{CRS}$, if the
distribution of its first output, i.e., $r$, is computationally
indistinguishable from $U_n$ (the uniform distribution over
strings of length $n$), and $\mathcal{R}_{CRS}(r, \tau_r)=1$ for
all outputs of $\mathcal{M}_{CRS}$ (typically, $\tau_r$ is some
trapdoor information about $r$). For a positive polynomial
$s(\cdot)$, we denote by $(\{r_1, r_2, \cdot, r_{s(n)}\},
\{\tau_{r_1}, \tau_{r_2}, \cdots, \tau_{r_{s(n)}}\}) \\
\longleftarrow \mathcal{M}^{s(n)}_{CRS}(1^n)$ the output of the
experiment of running
$\mathcal{M}_{CRS}(1^n)$ \emph{independently} $s(n)$-times,  
 where  for
any $i$, $1\leq i\leq s(n)$, $ (r_i, \tau_{r_i})$ denotes the
output of the $i$-th independent execution of $\mathcal{M}_{CRS}$.


\textbf{$\mathcal{M}_{CRS}$ trivially achievable distribution.}
 Let $G $ be a set of pairs of integers
$\{(i_1,j_1), (i_2, j_2), \cdots, (i_t, j_t)\}$,  where  $1\leq
i_1 < i_2 <\cdots <i_t \leq s(n)$ and $1 \leq j_1,j_2, \cdots, j_t
\leq s(n)$ are distinct integers, and  $0\leq t\leq s(n)$ such
that $G$ is defined to be the empty set when $t=0$. Let
$\mathcal{M}_{s,n,G}$ be the probability distribution over
$(\{0,1\}^n)^{2s(n)}$, obtained by first generating $2s(n)-t$
 $n$-bit strings $\{x_m, y_k | m \in \{1, 2, \cdots, s(n)\},
 k \in \{1, 2, \cdots, s(n)\}-\{j_1,j_2,\cdots, j_t\} \}$,
by 
 running $\mathcal{M}(1^n)$
independently $2s(n)-t$ times, and then defining $y_{j_d}=x_{i_d}$
for $1 \leq d \leq t$ and taking $(x_1,x_2,\cdots, x_{s(n)}, y_1,
y_2, \cdots,y_{s(n)})$ as the output. A probability distribution
over $(\{0,1\}^n)^{2s(n)}$ is called {\it $\mathcal{M}$-trivially
achievable}, if it is a convex combination of $U_{s,n,G}$ over all
$G$'s.


Now, we are ready for a formal definition of concurrently
non-malleable  coin-tossing (CNMCT) in the CMIM setting of the BPK
model.

\begin{definition}  [concurrently non-malleable coin-tossing CNMCT]\label{CNMCTDEF}
Let $\Pi=\langle L, R\rangle$ be a two-party protocol in the BPK
model, where $L=(L_{KEY}, L_{PROOF})$ and $R=(R_{KEY},
R_{PROOF})$. We say that $\Pi$ is a
  concurrently non-malleable coin-tossing protocol in the
BPK model w.r.t.  some
  key-validating relations $\mathcal{R}^L_{KEY}$ and  $\mathcal{R}^R_{KEY}$,   if for any PPT $s(n)$-CMIM adversary $\mathcal{A}$ in the BPK model
  there exists a probabilistic (expected)
polynomial-time
  algorithm $S=(S_{KEY}, S_{PROOF})$  such that, for any sufficiently large $n$, any auxiliary
  input  $z \in \{0, 1\}^*$, any PPT CRS-simulating algorithm $\mathcal{M}_{CRS}$ and
  any  polynomial-time computable (CRS-trapdoor validating) relation $\mathcal{R}_{CRS}$, and any
  polynomial-time computable (SK-independence distinguishing)  relation $\mathcal{R}$ (with components drawn from $\{0, 1\}^* \cup
\{\perp\}$), the following hold, in accordance with the experiment
\textbf{$\textsf{Expt}_{\textup{CNMCT}}(1^n,  z)$}  described
below (page \pageref{CNMCTexperiment}):


 \begin{table}[!h]
\begin{center}

\begin{tabular} {|c|}
 \hline \label{CNMCTexperiment}

 \textbf{$\textsf{Expt}_{\text{CNMCT}}(1^n, z)$}\\ 

\begin{minipage}[t] {5.7in} 
\vspace{0.1cm}

\textbf{Honest left-player key-generation:}

$(PK_L, SK_L)\longleftarrow L_{KEY}(1^n)$. Denote by
$\mathcal{K}_L$ the set of all legitimate public-keys generated by
$L_{KEY}(1^n)$. \emph{Note that the execution of  $L_{KEY}$ is
independent from the simulation below. In particular, only the
public-key $PK_L$ is passed on to the simulator. }
\\

\textbf{The simulator $S=(S_{KEY}, S_{PROOF})$:}

$(PK_R, SK_R, SK^{\prime}_R)\longleftarrow S_{KEY}(1^n)$, where
the distribution of $(PK_R, SK_R)$  is identical with that of the
output of the key-generation stage of the honest  right-player $R$
(i.e., $R_{KEY}$), $\mathcal{R}^R_{KEY}(PK_R,
SK_R)=\mathcal{R}^R_{KEY}(PK_R, SK^{\prime}_R)=1$ and the
distributions of $SK_R$ and
$SK^{\prime}_R$ are identical and \emph{independent}. 
 \\

 $(str, sta) \longleftarrow S_{PROOF}^{\mathcal{A}(1^n,\  \ PK_L, \ PK_R,\  z)}(1^n,  z,
 PK_L, PK_R,
 SK_R)$. That is,   on inputs $(1^n, z, PK_L, PK_R, SK_R)$ and with
oracle access to $\mathcal{A}(1^n,  PK_L, PK_R, z)$, the simulator
$S$ outputs a simulated transcript $str$ and some state
information $sta$. Denote by $R_L=\{R_L^{(1)}, R_L^{(2)}, \cdots,
R_L^{(s(n))}\}$ the set of outputs of the $s(n)$ left sessions in
$str$ and by $R_R=\{R_R^{(1)}, R_R^{(2)}, \cdots, R_R^{(s(n))}\}$
the set of outputs of the $s(n)$ right sessions in $str$. The
state information $sta$ consists, among others,  of two sub-sets
(of $s(n)$ components each): $sta_L=\{sta_L^{(1)}, sta_L^{(2)},
\cdots, sta_L^{(s(n))}\}$ and $sta_R=\{sta_R^{(1)}, sta_R^{(2)},
\cdots, sta_R^{(s(n))})\}$.
 \emph{Note that $S$ does not  know secret-key $SK_L$ of honest left player, that is,
 $S$ can emulate the honest left-player only from its public-key $PK_L$.} \\

For any  $z\in \{0, 1\}^*$, we denote by $S(1^n,  z)$ the random
variable  $str$ (in accordance with above processes of $L_{KEY}$,
$S_{KEY}$,   and $S_{PROOF}$). For any $z\in \{0, 1\}^*$, any
$PK_L \in \mathcal{K}_L$ and $(PK_R, SK_R)\in
\mathcal{R}^R_{KEY}$, we denote by $S(1^n, z, PK_L, PK_R, SK_R)$
the random variable $S(1^n, z)$ specific to $(PK_L, PK_R, SK_R)$.
\\

\end{minipage}
\\
\hline
\end{tabular}
\end{center}

\end{table}

\begin{itemize}
\item \textbf{\textup{ Simulatability.}}
The following ensembles are  indistinguishable: \\
 $\{S(1^n, z, PK_L, PK_R, SK_R)\}_{1^n, PK_L\in \mathcal{K}_L, (PK_R, SK_R)\in
 \mathcal{R}^R_{KEY}, z\in\{0, 1\}^*}$
 and

   $\{view^{L(SK_L), R(SK_R)}_{\mathcal{A}}(1^n,  z, PK_L, PK_R)\}_{1^n, PK_L\in \mathcal{K}_L, (PK_R, SK_R)\in
   \mathcal{R}^R_{KEY}, z\in\{0, 1\}^*}$
(defined in accordance with the   experiment
$\textsf{Expt}^{\mathcal{A}}_{CMIM}(1^n, z)$ depicted in Section
\ref{CMIMBPK}, page \pageref{CMIMexpt}). This in particular
implies that
 the probability
ensembles $\{S(1^n, z)\}_{1^n, z\in\{0, 1\}^*}$ and
$\{view_{\mathcal{A}}(1^n, z)\}_{1^n, z\in\{0, 1\}^*}$ are
indistinguishable.


%



\item \textbf{\textup{Strategy-restricted and predefinable
randomness.}} With overwhelming probability, the distributions of
$(R_L, sta_L)$ and $(R_R, sta_R)$ are  identical to that of
$\mathcal{M}^{s(n)}_{CRS} (1^n)$; furthermore, the distribution of
$(R_L, R_R)$ is $\mathcal{M}$-trivially achievable. 




\item \textbf{\textup{Secret-key independence.}}
$\Pr[\mathcal{R}(SK_R, str, sta)=1]$ is negligibly close
 to $\Pr[\mathcal{R}(SK^{\prime}_R,
str, sta)=1]$.

\end{itemize}

The probabilities are taken over the randomness of $S$ in the
key-generation stage (i.e., the randomness for generating $(PK_R,
SK_R, SK^{\prime}_R)$) and in all proof stages,   the randomness
of $L_{KEY}$, the randomness of  $\mathcal{M}_{CRS}$, and
the randomness of $\mathcal{A}$. 

\end{definition}

\subsubsection{Comments and
clarifications}\label{CNMCTclarification}

 Some comments and
clarifications  on the CNMCT definition are in place.


\textbf{On the strategy-restricted and predefinable randomness
property.} Note that the formalization of the strategy-restricted
and predefinable randomness property requires  that:  the
coin-tossing outputs of all left sessions (resp., all right
sessions) are \emph{independent} (pseudo)random strings and are
set
by the simulator $S$, in an online way at its wish.  
 We stress that we do
\emph{not} require, by such formalization, that the coin-tossing
outputs of \emph{all left and right} sessions are independent.
That is, we do not require the distribution of $((S_L, S_R)
(sta_L, sta_R))$ is identical to that of
$\mathcal{M}^{2s(n)}_{CRS}(1^n)$. The later formalization rules
out the natural copying strategy by definition, and thus is too
strong to naturally capture  CNM-secure cryptographic protocols. On
the other hand, in order to allow the copying strategy to the CMIM
adversary,  another alternative relaxed formalization is: we only
require that the coin-tossing output of each \emph{individual}
(left or right) session is identical to $\mathcal{M}_{CRS}(1^n)$.
But, this alternative formalization is too week to rule out
naturally insecure protocols. Specifically, consider that  a CMIM
adversary manages to set the outputs of some (and maybe all)
sessions to be the same string or to be maliciously correlated in
general. In this case, it is still can be true that the output of
each individual session is still identical to
$\mathcal{M}_{CRS}(1^n)$, but clearly not secure as coin-tossing
outputs are maliciously correlated.

\emph{Our formalization essentially implies that:  the
coin-tossing output of each left (resp.,  right)  session is either
  independent of  the outputs of all other sessions OR  copied
from the output of  one right (resp., left) session in another
CMIM part; furthermore, the output of each session in one CMIM
part can be copied into another CMIM part at most once.}



\textbf{On the ability of online setting all coin-tossing  outputs
and its implication of CNM security against \textsf{CMIM of full
adaptive input selection}.} Note that in the above CNMCT
formulation, the simulator $S$ not only outputs a simulated
transcript that is indistinguishable from the real view of the
CMIM adversary, but also, $S$ sets and controls, at the same time
\emph{in an online way},  the coin-tossing  outputs of \emph{all} left
and right sessions in the simulated transcript (in the sense that
$S$ knows the corresponding trapdoor information of \emph{all} the
coin-tossing outputs appearing in the simulated transcript). This
ability of $S$ plays several essential roles: Firstly, setting the
outputs of all CNMCT sessions (at its wish in an online way)  is
essential, in general, to transform CNM cryptography in the CRS
model into CNM cryptography in the BPK model, as in the security
formulation and analysis of CNM protocols in the CRS model the
simulator does control and set  all simulated CRS;  Secondly, such
ability of $S$ is critical for obtaining CNM security against
\textsf{CMIM with full adaptive input selection}, which is
addressed in detail below.

For more detailed clarifications about this issue, consider a
protocol (e.g., a ZK protocol) that is resulted from the
composition of a coin-tossing protocol in the BPK model and a
protocol (e.g., an NIZK protocol) in the CRS model, and assume the
CMIM adversary sets input to each session of the composed protocol
at the last message of that session. In particular,   the input to
each session can be an arbitrary function of  the coin-tossing
output and will be different with respect to  different
coin-tossing outputs. Now, suppose the simulator/extractor cannot
set the coin-tossing outputs of all  right sessions \emph{in an
online way};
 That is, for some (at least
one) successful right sessions in the simulated transcript, the
simulator fails in setting  the coin-tossing outputs of  these
sessions, and thus learning no trapdoor information enabling
on-line knowledge-extraction. In case  the inputs of these right
sessions do not appear as inputs of left sessions, then,
 in order  to extract witnesses to the inputs of such successful
right sessions appeared in the simulated transcript, the
simulator/extractor has to rewind the CMIM adversary and manages
to set,  one by one sequentially,  the  coin-tossing outputs  of
these right sessions. But, the problem is: whenever the
simulator/extractor is finally able to  set (if it is possible),
at its wish, the output of a right session in question, the input
to that right session set by the CMIM adversary is however changed
(as it is determined by the output of coin-tossing). This means
that the simulator/extractor may never be able to extract the
witnesses to all the inputs of successful right sessions appeared
in the simulated transcript.  The above arguments also apply to
Feige-Shamir-ZK-like protocols as illustrated in Section
\ref{fulladaptive}. We remark that, it is the ability of online
setting the outputs of all coin-tossing sessions, in our CNMCT
formulation and security analysis, that enables us to obtain
CNM
security against \textsf{CMIM of full adaptive input selection}. 

\textbf{On the generality of CNMCT.} We first note that CNMCT in
the BPK model  actually implies (or serves as the basis to formulate)  concurrent non-malleability with
full adaptive input selection for any cryptographic protocols in
the BPK model. The reason is: concurrent non-malleability for any
functionality can be implemented  in the common random string
model \cite{DDOPS01,CLOS02}. By composing any concurrent
non-malleable cryptographic protocol in the CRS model with a CNMCT
protocol in the BPK model, with the output of CNMCT serving as the
common random string of the underlying CNM-secure protocol in the
CRS model,  we can transform it into  a CNM-secure protocol in the
BPK model. In particular, we can view the composed protocol as a
special (extended) coin-tossing protocol. 
Specifically, to define the CNM security for any protocol in the
BPK model, which is resulted from the composition of a CNM-secure
protocol in the CRS model and a CNMCT protocol in the BPK model,
we just view the composed protocol as a special (extended)
coin-tossing protocol, and apply the CNMCT formulation to get the
CNM security formulation for the composed protocol. 

With CNMZK as an illustrative example, when composed with adaptive
non-malleable NIZK arguments of knowledge protocols (e.g., the
robust NIZK of \cite{DDOPS01} for $\mathcal{NP}$), CNMCT implies (tag-based\footnote{The tag-based CNM
security of the composed protocol is inherited from that of robust
NIZK. Here, we note that the CNM security formulation and protocol
implementation of robust NIZK \cite{DDOPS01} actually implies
tag-based CNM security, though it was not explicitly mentioned and
formalized there.})
concurrent non-malleable zero-knowledge arguments of knowledge (for $\mathcal{NP}$)
with full adaptive input selection in the BPK model. But, we do
not need to explicitly formulate the (adaptive input-selecting)
CNM security for ZK protocols in the BPK model.  Specifically, we
can view the composed protocol (of CNMCT and robust NIZK) as a
special version of coin-tossing and note that in this case $(str,
\tau)$ implies knowledge-extraction. Then, the properties of
simulatability and strategy-restricted and predetermined
randomness of CNMCT implies 
 simulation-extraction,  by viewing
$\mathcal{M}_{CRS}$ as the CRS simulator of the underlying NMNIZK.
The secret-key independent knowledge extraction is derived from
the property of secret-key independence of CNMCT.

\subsection{Implementation and analysis of constant-round CNMCT in the BPK model}\label{analysis}

\textbf{High-level overview of the CNMCT implementation.}
We design a coin-tossing
mechanism in the BPK model, which allows each player to set the
coin-tossing output whenever it learns its peers's secret-key. The
starting point is the basic and famous  Blum-Lindell coin-tossing
  \cite{B82,L01}:  the left-player $L$ commits a random
string $\sigma$, using randomness $s_{\sigma}$, to $c=C(\sigma,
s_{\sigma})$ with   a statistically-binding commitment scheme $C$;
  The right-player $R$ responds with a random string
$r_r$; $L$ sends back $r=\sigma \oplus r_l$ and proves the
knowledge of $(\sigma, s_{\sigma})$. To render the simulator the
ability of online setting coin-tossing outputs 
 against malicious right-players, $R$   proves its knowledge of its
secret-key $SK_R$ (using the key-pair trick of \cite{NY90}), and
$L$ accordingly  proves the knowledge of either $(\sigma,
s_{\sigma})$ or  $SK_R$. To render  the ability of online
setting coin-tossing outputs 
 against malicious left-players, $L$ registers $c=C(\sigma, s_{\sigma})$ as its public-key and treats
 $\sigma$ as the seed of a pseudorandom function PRF; $L$ then sends $r^{\prime}_l$  
  that commits  to $r_l=PRF_\sigma(r^{\prime}_l)$; after receiving
 $r_r$ from $R$, it returns back $r=r_l\oplus r_r$ and proves the
 knowledge of either its secret-key  $SK_L=(\sigma, s_{\sigma})$ (such that
 $r=r_r\oplus  PRF_\sigma(r^{\prime}_l)$)  or the right-player's
 secret-key $SK_R$.
  The underlying proof of knowledge is implemented with PRZK.
But, \emph{correct}   knowledge-extraction  with bare public-keys  in the complex CMIM setting is quite subtle. 
  At a very high level, the correct knowledge
extraction, as well as the CNM security, is reduced to the
one-left-many-right
non-malleability of PRZK. 

Now, we present the implementation of constant-round CNMCT
$\langle L, R\rangle$ in the BPK model, which is depicted in
Figure \ref{CNMCT} (page \pageref{CNMCT}). Each player
$L=(L_{KEY}, L_{PROOF})$ or $R=(R_{KEY}, R_{PROOF})$ works in two
stages: the key-generation stage (to be run by $L_{KEY}$ and
$R_{KEY}$) and the proof stage (to be run by $L_{PROOF}$ and
$R_{PROOF}$). But, for presentation simplicity, we often write $L$
and $R$ directly without explicitly indicating the key-generation
algorithm and the proof algorithm (which are implicitly clear from
the context).


\begin{center}
\begin{figure}[!t]
\begin{tabular} {|c|}
 \hline
\begin{minipage}[t]{6.4in} \small 

\begin{description}
\item [Right-player key registration:] Let $f:\{0,
1\}^*\rightarrow\{0, 1\}^*$ be a one-way function. On a security
parameter $n$, the right-player $R$ (actually $R_{KEY}$) randomly
selects $s_0, s_1$ from $\{0, 1\}^{n}$, computes $y_0=f(s_0)$,
$y_1=f(s_1)$. $R$ publishes  $PK_R=(y_0, y_1)$ as its public-key,
and keeps $SK_R=s_b$ as its secret-key for a random bit $b\in \{0,
1\}$  while discarding $SK^{\prime}=s_{1-b}$.
 Define $\mathcal{R}^R_{KEY}=\{((y_0, y_1), x)| y_0=f(x) \vee y_1=f(x)
 \}$, and $\mathcal{K}_R$ the corresponding $\mathcal{NP}$-language.

\item [Left-player  key registration:] Let $C$ be a
(non-interactive) statistically-binding commitment scheme. Each
left-player $L$ (actually $L_{KEY}$) selects $\sigma  \in \{0,
1\}^n$ and $s_{\sigma} \in \{0, 1\}^{poly(n)}$ uniformly at
random, computes $c=C(\sigma, s_{\sigma})$ (i.e., committing to
$\sigma$ using randomness $s_{\sigma}$). Set $PK_L=c$ and
$SK_L=(\sigma, s_{\sigma})$, where $\sigma$ serves as the random
seed of a pseudorandom function
$PRF$. 
 Define $\mathcal{K}_L=\{c| \exists (x, s) \ s.\ t. \ \  c=C(x, s)\}$. (We note that the left-player actually can also use Naor's OWF-based
statistically-binding commitment scheme, in this case each right
player's public-key will additionally include a $3n$-bit string
serving as the first-round of Naor's commitment scheme.)

\item [Note on fixed vs. interchangeable roles]  In the above
key-registration description, we have assumed protocol players do
not interchange their roles. This is critical for achieving CNM
security against CMIM adversary capable of full adaptive input
selection in the BPK model. But, as clarified in Section
\ref{CNMZKclarification}, each player can also choose the ability
of playing both (left-player and right-player) roles, by setting
the public-key to be $PK=(PK_L, PK_R)$ and the secret-key to be
$SK=(SK_L, SK_R)$. In this case, this player may lose CNM security
against adaptive input selecting CMIM adversary, but still hold
CNM security with predetermined inputs in the BPK model. That is,
whether playing with fixed role or interchangeable roles can be at
the discretion of each individual player. 
 The system may involve players of fixed role, 
 as well as players of
interchangeable role. 

\end{description}

\end{minipage}\\ \hline

\begin{minipage}[t]{6.4in} \small
\begin{description}
\item [Stage-1.] The right-player $R$ (actually $R_{PROOF}$)
computes  and sends $c_{sk}=C(SK_R, s_{sk})$,  where $C$ is a
constant-round statistically-binding commitment scheme and
$s_{sk}$ is the randomness used for commitment; Define
$\mathcal{L}_{SK}=\{((y_0, y_1),c_{sk})| \exists (s_{sk}, SK)  \
s.t. \ c_{sk}=C(SK, s_{sk})\wedge (y_0=f(SK) \vee y_1=f(SK))\}$.
Then, $R$ proves to the left-player $L$ the knowledge of $(SK_R,
s_{sk})$ such that $((PK_R, c_{sk}), (SK_R, c_{sk}))\in
\mathcal{R}_{\mathcal{L}_{SK}}$, by running the Pass-Rosen
non-malleable ZK (PRZK) for $\mathcal{NP}$ with the tag set to be
$(PK_L, PK_R=(y_0, y_1))$ that is referred to as the \emph{right}
tag. The composed protocol of statistically-binding commitments
and PRZK is called \emph{commit-then-PRZK}.

\item[Stage-2.]  The left player $L$ (actually $L_{PROOF}$)
randomly selects $r^{\prime}_l\leftarrow \{0,1\}^n$, and sends
$r^{\prime}_l$ to $R$.

\item[Stage-3.] The right player $R$ randomly selects
$r_r\leftarrow\{0,1\}^n$ and sends $r_r$ to the left player.

\item[Stage-4.] The left player computes
$r_l=PRF_{\sigma}(r^{\prime}_l)$ (where $\sigma$ is the random
seed of $PRF$ committed in $L$'s public-key $PK_L$), and sends
$r=r_l\oplus r_r$ to the right player.

\item [Stage-5.] $L$ computes and sends
$c_{crs}=C(\sigma||s_{\sigma}, s_{crs})$, where ``$||$" denotes
the operation of string concatenation. Define
$\mathcal{L}_{CRS}=\{(PK_L=C(\sigma, s_{\sigma}), PK_R=(y_0, y_1),
r^{\prime}_l,
 r_r, r, c_{crs})|\exists (x, s,  s_{crs}) \ s.t.\ c_{crs}=C(x||s, s_{crs})
\wedge [(PK_L=C(x, s) \wedge PRF_x(r^{\prime}_l)=r\oplus r_r) \vee
y_0=f(x) \vee y_1=f(x)]\}$.
Then, $L$ proves to $R$ the knowledge $(\sigma, s_{\sigma},
s_{crs})$ such that $((PK_L, PK_R, r^{\prime}_l,
 r_r, r, c_{crs}), (\sigma, s_{\sigma},   s_{crs})) \in \mathcal{R}_{\mathcal{L}_{CRS}}$,
 by running the PRZK for $\mathcal{NP}$ with
the tag set to be $(PK_L, r_r, r)$ that is referred to as the
\emph{left} tag.
 That is, $L$ proves to $R$ that
either the value committed in $c_{crs}$ is $SK_L=(\sigma,
s_{\sigma})$ such that $PRF_{\sigma}(r^{\prime}_l)=r\oplus r_r$ OR
the $n$-bit prefix of the committed value is the preimage of
either $y_0$ or $y_1$.  W.l.o.g., we can assume the left-tag $(PK_L,
r_r, r)$ and the right-tag $(PK_L, y_0, y_1)$ are of the same
length (the use of the session tags will be clear in the security analysis).

\end{description}
\end{minipage}\\

\begin{minipage}[t]{6.4in}\small
\vspace{0.3cm}
The result of the protocol is the string $r$. We will use the convention that if one of the parties aborts (or fails to provide a valid proof) then the other party determines the result of the protocol.\\
\end{minipage}
\\
\hline

\end{tabular}

\caption{\label{CNMCT} \small  Constant-round CNMCT in the BPK
model}

\end{figure}

\end{center}

\textbf{Notes on CNMCT implementation:} Note that  the PRZK is
used as a building tool in the coin-tossing protocol. That is,
PRZK is composed concurrently with other sub-protocols (rather
than composed concurrently with itself). Also
note that the tag of PRZK in Stage-5 is set interactively. 
For presentation simplicity, we have described commit-then-PRZK,
as well as PRZK, to work on concrete statements in Stage-1 and
Stage-5. In  actual implementation, both   commit-then-PRZK and
PRZK work for some $\mathcal{NP}$-Complete languages, and the
actual statements to be proved by   commit-then-PRZK and PRZK are
got by applying $\mathcal{NP}$-reductions, while the tags
remaining unchanged. With Stage-1 as the illustration example, the
verifier actually first reduces $PK_R$ into an instance, denoted
$s_{PK_R}$, of some $\mathcal{NP}$-Complete language, which serves
as the input to commit-then-PRZK of  Stage-1 and  the
statistically-binding commitment $c_{sk}$ actually commits to the
corresponding $\mathcal{NP}$-witness of $s_{PK_R}$; then, the
actual input to the subsequent PRZK is reduced from $(s_{PK_R},
c_{sk})$. The same treatment also applies to Stage-5. Note that
the left and right tag strings could be arbitrarily different from
(thought still polynomially related to)  the actually statements
reduced
by $\mathcal{NP}$-reductions.  
 We remark that in the actual implementation of
the above CNMCT protocol, PRZK can be replaced by any adaptive
tag-based one-left-many-right non-malleable (in the sense of
simulation-extraction) statistical ZK argument of knowledge for
$\mathcal{NP}$. But, PRZK is currently the only known one.

\begin{theorem}\label{CNMCTtheo}
Assuming OWF, and one-left-many-right adaptive tag-based non-malleable ZK
arguments of knowledge for $\mathcal{NP}$ (in the sense of
simulation/extraction),  the protocol $\Pi=\langle L, R\rangle$
depicted in Figure \ref{CNMCT} is a constant-round concurrent
non-malleable coin-tossing protocol in the BPK model.
\end{theorem}

\textbf{Proof (sketch).} \quad

\textbf{Underlying complexity assumptions}

Note that PRF can be implemented with any OWF \cite{GGM86,HILL99},
and the players can use Naor's OWF-based statistically-binding
commitments in key-registration. The (only) known adaptive
tag-based one-left-many-right
 non-malleable statistical  ZK argument of knowledge for $\mathcal{NP}$ is the
 Pass-Rosen ZK \cite{PR05s,PR05}, which is in turn based on
 collision-resistant hash function \cite{B01,BG02}.

 \textbf{The (high-level) description of the simulator}

On security parameter $1^n$, for   any positive polynomial
$s(\cdot)$ and any PPT $s(n)$-CMIM adversary $\mathcal{A}$ in  the
BPK model with auxiliary information $z\in \{0, 1\}^*$, the
simulator $S=(S_{KEY}, S_{PROOF})$, with respect to  the honest
left-player key-registration algorithm $L_{KEY}$ and  a CRS
simulating algorithm $\mathcal{M}_{CRS}$ is depicted in Figure
\ref{CNMCTsimu} (page \pageref{CNMCTsimu}). In the description,
the notation of $m$ denotes a message sent by the simulator
(emulating honest players), and $\tilde{m}$ denotes the arbitrary
message sent by the CMIM-adversary $\mathcal{A}$.

\begin{center}

\begin{figure}[!p]
\label{cnmsim}
\begin{tabular} {|l|l|} \hline
\multicolumn{2}{|l|}{\rule[-4mm]{0mm}{10mm}\parbox[t]{6.4in}{\small
\textbf{External honest left-player key-generation:} Let $(PK_L,
SK_L)\longleftarrow L_{KEY}(1^n)$, where $PK_L=c$ and
$SK_L=(\sigma, s_{\sigma})$ such that $\sigma\in \{0, 1\}^n$ and
$s_{\sigma}\in \{0, 1\}^{t(n)}$ and  $c=C(\sigma, s_{\sigma})$.
This captures the fact that $S$ does not know $SK_L$ and can
emulate the honest left-player with the same public-key $PK_L$.


}} \\
\hline

\multicolumn{2}{|l|}{\rule[-4mm]{0mm}{10mm}\parbox[t]{6.4in}{\small
\textbf{Public-key file generation:}

 $S_{KEY}(1^n)$   perfectly emulates the
key-generation stage of the honest right-player, getting
$PK_R=(y_0=f(s_0), y_1=f(s_1))$ and $SK_R=s_b$ and
$SK^{\prime}_R=s_{1-b}$ for a random bit $b$.

 Denote by $F^{\prime}$  the
list of at most $s(n)$ public-keys generated by $\mathcal{A}$ on
$(1^n, PK_L, PK_R, z)$, then  the public-key file of the system is
$F=F^{\prime}\cup \{PK_L, PK_R\}$ (i.e., the proof stages are
w.r.t. $F$). 
 }}


 \\ \hline

 \multicolumn{2}{|l|}{\rule[-4mm]{0mm}{10mm}\parbox[t]{6.4in}{\small

  $\mathcal{S}\leftarrow \{(PK_R, SK_R)\}$
 (i.e. initiate the set of covered keys $\mathcal{S}$  to be $\{(PK_R, SK_R)\}$).
 \\
  On input $(1^n, z, F^{\prime}, PK_L, PK_R, SK_R)$
  and with oracle access to $\mathcal{A}(PK_L, PK_R, F^{\prime}, z)$,
  the following process is run by $S_{PROOF}$ repeatedly at most $s(n)+1$ times.
  In each simulation repetition, $S$ tries to  either end with a successful simulation or
  cover a new public-key in $F-\mathcal{S}$.\\  }}
 \\ 


 \parbox[t]{3.0in}{\small   \textbf{Straight-line left
simulation: }  \\
 In the $i$-th left concurrent session
 (ordered by the time-step in which the first round of each session is played)
 between $S$ and $\mathcal{A}$ in the left CMIM interaction part with respect to
  a public-key $PK^{(j)}_R=(y^{(j)}_0, y^{(j)}_1)\in \mathcal{K}_R$, $1\leq i, j\leq s(n)$,   $S$  acts as follows:
  \\
  \\
In case $\mathcal{A}$ successfully finishes Stage-1
 and  $PK^{(j)}_R\in
 F^{\prime}-\mathcal{S}$, the simulator ends the current repetition of simulation trial,  and
 starts to extract a secret-key $SK^{(j)}_R$ such that
 $\mathcal{R}^R_{KEY}(PK^{(j)}_R, SK^{(j)}_R)=1$, which is guaranteed by the AOK property of PRZK.
 Then, let
 $\mathcal{S}\leftarrow \mathcal{S}\cup\{(PK^{(j)}_R, SK^{(j)}_R)\}$, and move to
 next repetition with fresh randomness (but with the accumulated covered-key set $\mathcal{S}$ and the same
 public-key file $F$).
 \\
\\
In case $\mathcal{A}$ successfully finishes Stage-1
 and  $PK^{(j)}_R\in
 \mathcal{S}$ (i.e., $S$ has already learnt the secret-key $SK^{(j)}_R$),
   $S$ randomly  selects $r^{(i)\prime}_l\leftarrow\{0,1\}^n$ and
   sends  $r^{(i)\prime}_l$  to $\mathcal{A}$ at
Stage-2. After receiving Stage-3 message, denoted
$\tilde{r}_r^{(i)}$,  from $\mathcal{A}$, $S$ invokes
$\mathcal{M}_{CRS}(1^n)$ and gets the  output  denoted
$(S_L^{(i)}, \tau_L^{(i)})$. $S$ then sends $r^{(i)}=S_L^{(i)}$ as
the Stage-4 message (rather than sending back
$r^{(i)}=PRF_{\sigma}(r^{(i)\prime}_l) \oplus \tilde{r}_r^{(i)}$
as the honest left-player does), and sets
$sta_L^{(i)}=\tau_L^{(i)}$. In Stage-5, $S$ computes and sends
$c^{(i)}_{crs}=C(SK^{(j)}_R||0^{t(n)}, s^{(i)}_{crs})$ to
$\mathcal{A}$ (rather than sending back
$c^{(i)}_{crs}=C(\sigma||s_{\sigma})$ as the honest left-player
does), where $t(n)$  is the length of $s_{\sigma}$ in $SK_L$.
Finally, $S$ finishes the PRZK of Stage-5 with $(SK^{(j)}_R,
s^{(i)}_{crs})$ as its witness and $(PK_L, \tilde{r}_r^{(i)},
S^{(i)}_L)$
as the tag.

} &
\parbox[t]{3.4in}{\small \textbf{Straight-line  right
simulation:} \\ In the $i$-th right concurrent session
 (ordered by the time-step in which the first round of each session is played)
 between $S$ and $\mathcal{A}$ in the right CMIM interaction part with respect to
  a public-key $PK^{(j)}_L=c^{(j)}\in \mathcal{K}_L$, $1\leq i, j\leq s(n)$,   $S$  acts as follows:
  \\

$S$ perfectly emulates honest right-player in Stage-1 of any right
session, with $SK_R$ as the witness to commit-then-PRZK and
$(PK^{(j)}_L, PK_R)$ as the tag.
\\

\textbf{Case-R1:} If $PK^{(j)}_L\in
 \mathcal{S}$ (i.e., $S$ has already learnt the secret-key $SK^{(j)}_L=(\sigma^{(j)},
 s^{(j)}_{\sigma})$), after receiving $\tilde{r}^{(i)\prime}_l$ from $\mathcal{A}$  at
 Stage-2,  $S$ runs
$\mathcal{M}_{CRS}(1^n)$ and gets the output denoted  $(S_R^{(i)},
\tau_R^{(i)})$, and then  computes and sends
 $PRF_{\sigma^{(j)}}(\tilde{r}^{(i)\prime}_l)\oplus S^{(i)}_R$  as Stage-3
 message, and goes further. 
 \\

\textbf{Case-R2:} If $PK^{(j)}_L\not\in
 \mathcal{S}\cup\{PK_L\}$, and $\mathcal{A}$ successfully finishes the $i$-th
 right session (in which $S$ just perfectly emulates the honest right-player of $PK_R$), 
then  the simulator $S$ ends the current repetition of simulation
trial,  and
 starts to extract a secret-key $SK^{(j)}_L$ such that
 $\mathcal{R}^L_{KEY}(PK^{(j)}_L, SK^{(j)}_L)=1$. \emph{In case $S$ fails to extract
 such $SK^{(j)}_L$, $S$ stops the simulation, and outputs a special
 symbol $\bot$ indicating simulation failure. Such simulation failure is called Case-R2 failure.} In case $S$ successfully  extracts
 such $SK^{(j)}_L$, then let
 $\mathcal{S}\leftarrow \mathcal{S}\cup\{(PK^{(j)}_L, SK^{(j)}_L)\}$, and move to
 next repetition with fresh randomness (but with the accumulated covered-key set $\mathcal{S}$
  and the same
 public-key file).
\\

\textbf{Setting $sta_R$:} For successful $i$-th right session, if
the Stage-4 message $\tilde{r}^{(i)}$ is $S^{(i)}_R$ or
$S^{(k)}_L$ for some  $k$, $1\leq k\leq s(n)$, then $sta^{(i)}_R$
is set accordingly to $\tau^{(i)}_R$ or $\tau^{(k)}_L$; otherwise,
$sta^{(i)}_R$ is set to be $\bot$.

 }
\\ \hline




\end{tabular}
\caption{\label{CNMCTsimu}  The CNM simulation} 
\end{figure}
\end{center}

\textbf{Notes on the CNM simulation:}  For any $i$, $1\leq i\leq
 s(n)$, if in the $i$-th left (resp., right) session of   the simulation $\mathcal{A}$ does not act
 accordingly or fails to provide a valid proof, then $S$ aborts that
 session, and sets the output  just to be $S^{(i)}_L$ (resp., $S^{(i)}_R$) and the state information to
 be $\tau^{(i)}_L$ (resp., $\tau^{(i)}_R$).

Note that in Case-R2 of right-session simulation (i.e., a successful right-session w.r.t. a left-player key $PK^{(j)}_L=PK_L$), the simulator does not try to extract the secret-key of $PK_L$.  In the following analysis, we show that in this case, with overwhelming probability,  the tag of Stage-5 of this successful right session  is identical to that of Stage-5 of a left-session. As the tag of Stage-5 of a session consists of the session output (i.e., the coin-tossing output), this implies that the session output of the right-session is identical to that of one of left-sessions. Moreover, we show that with overwhelming probability each left-session output can appear, as session output, in at most one successful right-session.


 In the unlikely event that $\mathcal{A}$ finishes a right session and the Stage-1 of a
 left-session simultaneously, both of which are w.r.t. uncovered public-keys, extracting
 $SK_R$ in left simulation part  takes priority (in this case, $SK_L$ extraction in right
 simulation part  is ignored in the current simulation repetition).

 During any (of the at most $s(n)+1$)  simulation repetition, if $S$ does not encounter
 secret-key extraction and does not stop due to Case-R1 failure or Case-R2 failure,
  then  $S$ stops whenever $\mathcal{A}$
 stops, and sets $str$ to be   $F$ and the view of $\mathcal{A}$ in
 this simulation repetition and $sta=(sta_L, sta_R)$ to be the according
 state-information.

\textbf{Analysis of the CNM simulation}

In order to establish the CNM security of the coin-tossing
protocol depicted in Figure \ref{CNMCT}, according to the CNMCT
definition of Definition \ref{CNMCTDEF}, we need to show the
following properties of the CNM simulator $S$ described in Figure
\ref{CNMCTsimu}:
\begin{itemize}
\item \textsf{$S$ works in expected polynomial-time.}

\item \textsf{The simulatability property, i.e., the output of $S$
is computationally indistinguishable from the view of
$\mathcal{A}$ in real CMIM attack.} \item \textsf{The property of
strategy-restricted and predefinable randomness.} \item
\textsf{The secret-key independence property.}

\end{itemize}

In the following, we analyze the above four properties of the CNM
simulator $S$ case by case.

\begin{itemize}
\item \textsf{$S$ works in expected polynomial-time}
\end{itemize}

Note that $S$ works for  at most $s(n)+1$ repetitions. Then,
pending on the ability of $S$ to extract secret-key of uncovered
public-keys in expected polynomial-time during each repetition
(equivalently, within running-time inversely propositional to the
probability of secret-key extraction event occurs), $S$ will work
in expected polynomial-time. The technique for covering
public-keys follows that
of \cite{CGGM00,BGGL01}. 
 Below,  we specify the secret-key extraction procedures in
more details.

\textbf{Right-player key coverage.} Whenever $S$  needs to extract
the secret-key $SK^{(j)}_R$ corresponding to an uncovered
public-key $PK^{(j)}_R$, due to successful Stage-1 of  the $i$-th
left session during the $k$-th simulation repetition w.r.t.
covered key set $\mathcal{S}^{(k)}$, $1\leq i, j \leq s(n)$ and
$1\leq k \leq s(n)+1$, we combine the CMIM adversary $\mathcal{A}$
and the simulation other than Stage-1 of the $i$-th left session
(i.e., the public file $F$, the covered key set
$\mathcal{S}^{(k)}$, the randomness $r_{\mathcal{A}}$ of
$\mathcal{A}$,  and the randomness $r_{\mathcal{S}}$ used by $S$
except for that to be used in Stage-1 of the $i$-th left session)
into an imaginary (deterministic) knowledge prover $\hat{P}^{(i,
j)}_{(\mathcal{S}^{(k)},r_\mathcal{A}, r_\mathcal{S})}$. Note
that, by the description of the CNM simulation depicted in Figure
\ref{CNMCTsimu},  the Stage-1 of the $i$-th left session is the
\emph{first successful} Stage-1 of a left session finished by
$\mathcal{A}$ (during the $k$-th simulation repetition) with
respect to an uncovered public-key not in $\mathcal{S}^{(k)}$. The
knowledge-prover $\hat{P}^{(i,
j)}_{(\mathcal{S}^{(k)},r_\mathcal{A}, r_\mathcal{S})}$ only
interacts with a stand-alone knowledge-verifier of
commit-then-PRZK, by running $\mathcal{A}$ internally and
mimicking $S$ with respect to $\mathcal{S}^{(k)}$ but with the
following exceptions: (1) the messages belonging to the Stage-1 of
the $i$-th left session are relayed  between the internal
$\mathcal{A}$ and the external stand-alone knowledge-verifier of
PRZK; (2) $\hat{P}^{(i, j)}_{(\mathcal{S}^{(k)},r_\mathcal{A},
r_\mathcal{S})}$ ignores the events of secret-key extraction in
right simulation part, i.e., successful right sessions with
respect to uncovered (left-player) public-keys; (3)  whenever
$\mathcal{A}$ (run internally by $\hat{P}^{(i,
j)}_{(\mathcal{S}^{(k)},r_\mathcal{A}, r_\mathcal{S})}$)
successfully  finishes, \emph{for the first time}, Stage-1 of a
left session w.r.t. an uncovered (right-player) public-key not in
$\mathcal{S}^{(k)}$, $\hat{P}^{(i,
j)}_{(\mathcal{S}^{(k)},r_\mathcal{A}, r_\mathcal{S})}$ just
stops.

For any intermediate $\mathcal{S}^{(k)}$ used in the $k$-th
simulation repetition,  any $PK^{(j)}_R\not\in \mathcal{S}^{(k)}$,
any  randomness $r_{\mathcal{A}}$ of $\mathcal{A}$ and any
randomness $r_{\mathcal{S}}$ used by $S$ except for that to be
used in Stage-1 of the $i$-th left session,   denote by $p$ the
probability (taken over the coins used by $\mathcal{S}$ for
Stage-1 of the $i$-th
left session)  
 that the public-key used by $\mathcal{A}$ in Stage-1 of the $i$-th left session  is $PK^{(j)}_R$,
 and furthermore,
  the Stage-1 of $i$-th left session is the first \emph{fist} successful
Stage-1 of a left session w.r.t. an uncovered public-key  during
the simulation of $\mathcal{S}$ w.r.t. covered-key set
$\mathcal{S}^{(k)}$. In other words, $p$  is the probability,
taken over the coins used by $\mathcal{S}$ for Stage-1 of the
$i$-th left session (but for fixed other coins), 
 of the event that  $\mathcal{S}$ needs to cover
$PK^{(j)}_R\not\in \mathcal{S}^{(k)}$ in the $i$-th left session
in its simulation w.r.t. $\mathcal{S}^{(k)}$.
 Clearly, with
probability at least $p$, the knowledge prover  $\hat{P}^{(i,
j)}_{(\mathcal{S}^{(k)},r_\mathcal{A}, r_\mathcal{S})}$
successfully convinces the stand-alone knowledge verifier of
$PK^{(j)}_R$. By the AOK property of PRZK and applying the
knowledge-extractor   on
$\hat{P}^{(i,j)}_{(\mathcal{S}^{(k)},r_\mathcal{A},
r_\mathcal{S})}$, the secret-key $SK^{(j)}_R$ will be extracted
within running-time inversely propositional to $p$.
 Here, when $p$  is
negligible, standard technique, originally proposed in \cite{GK96}
and then deliberated
 in \cite{L01},  has to be applied here (to estimate the value
of $p$) to make sure expected polynomial-time
knowledge-extraction. In more detail, the running-time of the
naive approach to directly applying knowledge-extractor whenever
such events occur is bounded by $T(n)=p \cdot
\frac{q(n)}{p-\kappa(n)}$, where $\kappa(n)$ is the
knowledge-error and $q(\cdot)$ is the polynomial related to the
running time of the knowledge-extractor that is
$\frac{q(n)}{p-\kappa(n)}$. The subtle point is: when $p$ is
negligible, $T(n)$ is not necessarily to be polynomial in $n$. The
reader is referred to \cite{GK96,L01} for the technical details of
dealing with this issue.


\textbf{Left-player key coverage.}

The coverage procedure for uncovered (left-player) public-keys
used by $\mathcal{A}$ in successful Stage-5 of right sessions can
be described accordingly, similar to above right-player key
coverage. The key point to note here is: for a successful right
session with respect to an uncovered (left-player) public-key
$PK^{(j)}_L$,  the value extracted in expected polynomial-time is
not necessarily to be the secret-key $SK^{(j)}_L$, though the
value extracted must be either $SK^{(j)}_L$ or $SK_R$ (i.e., the
preimage of either $y_0$ or $y_1$) , where $PK_R=(y_0, y_1)$ is
the simulated (right-player) public-key. That is, $S$ may abort
due to Case-R2 failure (though it works in expected
polynomial-time). We show, in the following analysis of the
simulatability property, Case-R2 failure occurs with at most
negligible probability.

\begin{itemize}
\item \textsf{Simulatability}
\end{itemize}

For presentation simplicity, in the following  analysis of
simulatability  we assume the first output of $\mathcal{M}_{CRS}$
is truly random string of length $n$, i.e., all $S^{(i)}_L$'s and
$S^{(i)}_R$'s are truly random strings.   The extension of the
simulatability  analysis to the case of pseudorandom output of
$\mathcal{M}_{CRS}$ is direct.

Assuming truly random output of $\mathcal{M}_{CRS}$,  there are
three  differences between the simulated transcript output by $S$
and the view of $\mathcal{A}$ in real CMIM attack against the
honest left-player of $PK_L$ and the honest right-player of
$PK_R$:

\begin{description}
\item [Truly random vs. pseudorandom Stage-4 messages:]
 In simulation,  the simulator $S$ sends truly
random string $r^{(i)}=S^{(i)}_L$ at Stage-4 of the $i$-th left
session, for any $i$, $1\leq i \leq s(n)$. But, the honest
left-player sends a pseudorandom Stage-4 message, i.e.,
$r^{(i)}=PRF_{\sigma}(r^{(i)\prime}_l)\oplus \tilde{r}^{(i)}_r$,
where $r^{(i)\prime}_l$ and $\tilde{r}^{(i)}_r$ are the Stage-2
and Stage-3 messages of the $i$-th left session.

\item [Witness difference of Stage-5 of left sessions:] For any
$i$-th left session w.r.t. a public-key $PK^{(j)}_R\in
\mathcal{S}$, the witness used by $S$ in the commit-then-PRZK of
Stage-5 is always the extracted secret-key $SK^{(j)}_R$, while the
witness used by the honest left-player is always its secret-key
$SK_L$.


\item [Case-R2 failure:] $S$ may stop with simulation failure, due
to invalid secret-key extraction in Case-R2 in the right
simulation part.

\end{description}

We first show that,  conditioned on Case-R2 failure does not
occur, the output of $S$ is indistinguishable from the real view
of $\mathcal{A}$. Specifically, we have the following lemma:

\begin{lemma}\label {simuR2} Conditioned on Case-R2 failure does not  occur,
 the following ensembles are indistinguishable:
 $\{S(1^n, z, PK_L, PK_R, SK_R)\}_{1^n, PK_L\in \mathcal{K}_L, (PK_R, SK_R)\in
 \mathcal{R}^R_{KEY}, z\in\{0, 1\}^*}$ (defined in Definition
 \ref{CNMCTDEF})  and
   $\{view^{L(SK_L), R(SK_R)}_{\mathcal{A}}(1^n,  z, PK_L, PK_R)\}_{1^n, PK_L\in \mathcal{K}_L, (PK_R, SK_R)\in
   \mathcal{R}^R_{KEY}, z\in\{0, 1\}^*}$
(defined in accordance with the   experiment
$\textsf{Expt}^{\mathcal{A}}_{CMIM}(1^n, z)$ depicted in Section
\ref{CMIMBPK}, page \pageref{CMIMexpt}).
\end{lemma}

\textbf{Proof} (of Lemma \ref{simuR2}). We first note that,
conditioned on  Case-R2 failure does not  occur and assuming the
truly random output of $\mathcal{M}_{CRS}$, $S$ perfectly
emulates 
 the honest right-player of $PK_R$ in right simulation part.

The left two differences all are w.r.t. left session simulation.
Intuitively, in real interaction the seed $\sigma$ of $PRF$ is
committed  into left-player public-key $PK_L$  and is re-committed
and proved  concurrently in Stage-5 of left sessions, the CMIM
adversary may potentially gain  some knowledge  about the random
seed $\sigma$ by concurrent interaction, which enabling it to set
its Stage-3 messages of left sessions maliciously depending  on
the
output of $PRF_{\sigma}$.   
 Note that in real interaction, the Stage-4
messages sent by honest left-player are determined by the PRF seed
and the Stage-2 messages. Thus, the Stage-4 messages of left
sessions in real interaction may be distinguishable from truly
random strings as sent by the simulator $S$ in simulation. The
still indistinguishability between the simulated transcript and
the real view of $\mathcal{A}$ is proved by hybrid arguments.

We consider a hybrid mental experiment $\mathcal{H}$.
$\mathcal{H}$
mimics  
 $S(1^n, z, PK_L, PK_R, SK_R)$, with
additionally possessing $SK_L=(\sigma, s_{\sigma})$ and with the
following exception: At Stage-4 of any left session, $\mathcal{H}$
just emulates the honest left-player by  setting the Stage-4
message $r^{(i)}$  to be $PRF_{\sigma}(r^{(i)\prime}_l)\oplus
\tilde{r}^{(i)}_r$ (rather than sending $S^{(i)}_L$ as $S$ does);
In Stage-5 of any left session w.r.t. a covered key $PK^{(j)}_R$
(for which $\mathcal{H}$ has already learnt the corresponding
secret-key $SK^{(j)}_L$), $\mathcal{H}$ still emulates $S$ by
using the extracted secret-key $SK^{(j)}_R$ as the witness
(specifically, it commits to $SK^{(j)}_R||0^t$ and finishes PRZK
accordingly as the simulator $S$ does).

The difference between the view of $\mathcal{A}$ in $\mathcal{H}$
and the view of $\mathcal{A}$ in the simulation of $S$ lies in the
difference of Stage-4 messages of left sessions. Suppose that the
view of $\mathcal{A}$ in $\mathcal{H}$ is distinguishable from the
view of $\mathcal{A}$ in the simulation of $S$, then it implies
that there exists a PPT algorithm $D$ that, given the commitment
 of the PRF seed, i.e.,  $PK_L=C(\sigma, s_{\sigma})$, can
 distinguish the output of $PRF_{\sigma}$  from truly random
 strings. Specifically, 
 on input $PK_L$, $D$  emulates
 $\mathcal{H}$ or $S$ by having oracle access to $PRF_{\sigma}$
 or a truly random function; Whenever it needs to send Stage-4
 message in a left session, it just queries  its oracle with the
 Stage-2 message. Clearly, if the oracle is $PRF_{\sigma}$, then
 $D$ perfectly emulates $\mathcal{H}$, otherwise (i.e., the oracle
 is a truly random function), it perfectly emulates the simulation
 of $S$.

 So, we conclude that if  the
view of $\mathcal{A}$ in $\mathcal{H}$ is distinguishable from the
view of $\mathcal{A}$ in the simulation of $S$, then the PPT
algorithm $D$ that, given the commitment
 of the PRF seed $\sigma$, can
 distinguish the output of $PRF_{\sigma}$  from that of truly random function.  Consider the case that $D$,
 given the commitment $c=C(\sigma)$, has oracle access to an
 independent  $PRF_{\sigma^{\prime}}$ of an independent random seed
 $\sigma^{\prime}$ or a truly random function. Due to the pseudorandomness of $PRF$, the output of $D(c)$
 with  oracle access to  $PRF_{\sigma^{\prime}}$ is indistinguishable from  the output of $D(c)$
 with oracle access to   a truly random function. It implies that $D$, given the commitment $c=C(\sigma)$,  can
 distinguish the output of $PRF_{\sigma}$ and the output of
 $PRF_{\sigma^{\prime}}$, where $\sigma$ and $\sigma^{\prime}$ are
 independent random seeds. But, this violates the computational hiding
 property of the commitment scheme $C$. Specifically, given two
 random strings of length $n$,  $(s_0, s_1)$, and a commitment
 $c_b=C(s_b)$ for a random bit $b$, the algorithm $D$ can be used to
 distinguish the value committed in $c_b$, which violates the
 computational hiding property of $C$.

Now, we consider the difference between the output of
$\mathcal{H}$ and the view of $\mathcal{A}$ in real execution.
Recall that, as we have shown the view of $\mathcal{A}$ in
$\mathcal{H}$ is indistinguishable from that in the simulation and
we have assumed Case-R2 failure does not occur in the simulation
of $S$, Case-R2 failure can occur in
$\mathcal{H}$ with at most negligible probability. 
 Then, the difference  between the output of
 $\mathcal{H}$ and the view of
$\mathcal{A}$ in real execution  lies in the witnesses used in
Stage-5 of left sessions. Specifically, $\mathcal{H}$ still uses
the extracted right-player secret-keys in Stage-5 of left
sessions, while the honest left-player always uses its secret-key
$SK_L$ in Stage-5 of left sessions in real execution. By hybrid
arguments, the difference can be reduced to violate the regular WI
property of commit-then-PRZK. Note that commit-then-PRZK is itself
regular WI for $\mathcal{NP}$ (actually, any commit-then-SWI is
itself regular WI).

In more detail, we consider  the mental experiment $M_b$,
$b\in\{0, 1\}$. On input $\{(PK_L, SK_L), (PK_R, SK_R)\}$ and
public file $F$, and auxiliary information $z$ to the CMIM
adversary $\mathcal{A}$ \footnote{Recall that, in accordance with
the definition of CNMCT, $z$ is a priori information of
$\mathcal{A}$ that is independent from the public file $F$ (in
particular, $PK_L$ and $PK_R$).}, the mental $M_b$ also takes as
input all secret-keys corresponding to right-player public-keys in
the public file $F$ (in case the corresponding secret-keys exist).
$M_b$ runs the  CMIM adversary $\mathcal{A}$ as follows:

\begin{enumerate}

\item $M_b$ emulates the honest right-player of $PK_R$ (with
$SK_R$ as the witness)  in right sessions. In particular,  $M$
just sends truly random Stage-3 messages in all right sessions,
and ignores knowledge-extraction of left-player secret-keys in
right sessions (i.e., in case $\mathcal{A}$ successfully finishes
a right session w.r.t an uncovered public-key $PK^{(j)}_L$, $M_b$
ignores the need of secret-key extraction and just moves on);

\item  For any $i, j$, $1\leq i\leq s(n)$ and $1\leq j\leq
s(n)+1$, in the $i$-th left session w.r.t. right-player public-key
$PK^{(j)}_R$, $M_b$ emulates the honest left-player of $PK_L$
until Stage-4  (in particular, it sets the Stage-4 message
$r^{(i)}$  to be $PRF_{\sigma}(r^{(i)\prime}_l)\oplus
\tilde{r}^{(i)}_r$), but with the following exception in Stage-5:
\begin{itemize}
\item If $b=0$, then $M_b$ just emulates the honest left-player in
Stage-5 of the left session, with $SK_L$ as its witness.

\item If $b=1$,  $M_b$ still emulates the simulator  by using the
secret-key $SK^{(j)}_R$, for which we assume it exists and $M$
knows, as the witness in Stage-5. Specifically, it commits to
$SK^{(j)}_R||0^t$ and finishes PRZK accordingly as the simulator
$S$ does.
\end{itemize}

\end{enumerate}

It's easy to see that the output of  $M_0$ is identical to the
real view of $\mathcal{A}$ in real execution, and the output of
$M_1$ is indistinguishable from the output of $\mathcal{H}$. Then,
suppose the real view of $\mathcal{A}$ in real execution is
distinguishable from  the output of $\mathcal{H}$, by hybrid
arguments we can break the regular WI of commit-then-PRZK. \hfill
$\square$


 Now, we  show that Case-R2
 failure indeed
occurs with negligible probability, from which the simulatability
of the CNM simulation is established.

\begin{lemma}\label{R2}
Case-R2 failure occurs with negligible probability.
\end{lemma}

\textbf{Proof} (of Lemma \ref{R2}). Suppose Case-R2 failure occurs
with non-negligible probability. That
is, for 
some  polynomial $p(n)$ and  infinitely many $n$'s, with
probability of $\frac{1}{p(n)}$ there exist  $k, i, j$, $1\leq k
\leq s(n)+1$ and $1\leq i, j \leq s(n)$, such that in the $k$-th
simulation
repetition   
 $\mathcal{A}$ successfully finishes the $i$-th
right session  with respect to an uncovered public-key
$PK^{(j)}_L\not\in \mathcal{S}\cup \{PK_L\}$, 
 furthermore, the $k$-th
simulation repetition is the \emph{first} one encountering Case-R2
failure and the $i$-th right session is the \emph{first}
successful session w.r.t. an uncovered public-key not in
$\mathcal{S}\cup\{PK_L\}$ during the $k$-th simulation repetition, 
 but the simulator fails in
extracting the corresponding secret-key $SK^{(j)}_L$. Recall that
$S$ makes at most $s(n)+1$ simulation trials (repetitions) and
each simulation trial uses fresh randomness in the proof stages;
$S$ starts knowledge-extraction whenever it encounters a
successful session w.r.t. an uncovered public-key different from
$PK_L$; Whenever Case-R2 failure occurs $S$ aborts the whole
simulation, which implies that the $k$-th simulation repetition is
also the \emph{last} simulation trial.

Note that, \emph{by the AOK property of PRZK}  (we can combine the
$k$-th simulation repetition except for the Stage-5 of the $i$-th
right session into a stand-alone knowledge prover of the PRZK), in
this case the simulator still extracts some value that is uniquely
determined by the statistically-binding commitment
$\tilde{c}^{(i)}_{crs}$ at the start of Stage-5 of the $i$-th
right session. According to the AOK property of PRZK, there are
two possibilities for the value
committed to $\tilde{c}^{(i)}_{crs}$ and extracted by $S$ assuming Case-R2 failure. 


\begin{description}

\item [Case-1.] The value committed is the preimage of $y_{1-b}$.
Recall that $PK_R=(y_0, y_1)$ is the simulated public-key of
honest right player, with $SK_R=s_b$ for a random bit $b$ such
that $y_b=f(s_b)$.

\item [Case-2.] The value committed is the preimage of $y_b$.

\end{description}

Due to the one-wayness of the OWF $f$, it is easy to see that
Case-1 can occur only with negligible probability. Specifically,
consider the case that $y_{1-b}$ is given to the simulator, rather
than generated by the simulator itself.

Below,  we show that Case-2 occurs also with negligible
probability, from which Lemma \ref{R2} is then established.


 We consider the following two experiments: 
$E(1^n, s_b)$, where $b\in \{0, 1\}$.
 The experiment $E(1^n, s_b)$ consists of two phases, denoted by $E_1$
 and $E_2$:
  In the first
 phase, $E_1$ just runs  $S(1^n, s_b)$ until $S$ stops. Denote by $\mathcal{C}_b$ the
  set of extracted-keys, corresponding to public-keys in $F-\{PK_R\}$,  
which are  extracted and used  by $S(1^n, s_b)$  
 in its \emph{last} simulation trial (recall that the \emph{first}
simulation repetition encountering Case-R2 failure is also the
last simulation repetition). Specifically, suppose $S$ uses
$SK_R=s_b$ in the simulation and  stops in the $k$-th simulation
repetition with respect to covered-key set, denoted
$\mathcal{S}^{(k)}_b$, then
$\mathcal{C}_b=\mathcal{S}^{(k)}_b-\{(PK_R, SK_R)\}$. Note that
$\mathcal{C}_b$ does not
  include $(PK_R, SK_R)$ now.  The set $\mathcal{C}_b$ generated by $E_1$
is passed on to $E_2$.

Then, in the second phase of the experiment $E(1^n, s_b)$,
$E_2(1^n, s_b, \mathcal{C}_b)$ runs the CMIM adversary
$\mathcal{A}$ and (re)mimics the  simulation of $S$ at its last
simulation trial w.r.t. the
set of covered-keys $\mathcal{C}_b$,  but with the following exceptions: 
 (1) $E_2$ sends truly random Stage-3 message in each right session; (2) $E_2$   has oracle access
 to the prover of commit-then-PRZK $P(1^n, s_b)$;  Whenever $S$ needs
 to give a Stage-1 proof of a right session on $PK_R=(y_0, y_1)$, or
 needs to give  a Stage-5 proof of a left session with respect to  $PK_R$ \footnote{Note
 that left sessions
 may be with respect to the simulated public-key
 $PK_R$, i.e., the CMIM adversary may impersonate the honest right-player of $PK_R$ in left sessions.}
   on input $(PK_L, PK_R,
 (r^{(i)\prime}_l, \tilde{r}^{(i)}_r, r^{(i)}))$, $E_2$ just sets the
 corresponding input,  i.e., $PK_R$ or $(PK_L, PK_R,
 (r^{(i)\prime}_l, \tilde{r}^{(i)}_r, r^{(i)}_l))$ \footnote{Actually, the $\mathcal{NP}$-statements
 reduced from them for the $\mathcal{NP}$-Complete
 language for which commit-then-PRZK actually works.}, as well as the according left or right tag,
   to its oracle $P(s_b)$, and
 then  relays messages between the oracle and the CMIM adversary
 $\mathcal{A}$; (3) In case $\mathcal{A}$ successfully finishes Stage-1
 of a left session or Stage-5 of a right session with respect to
 an uncovered public-key not in $\mathcal{C}_b \cup
 \{PK_L, PK_R\}$ in the run of $E_2(1^n, s_b, \mathcal{C}_b)$, $E_2$ just
 stops.


Now,  suppose Case-2 of Case-R2 failure  occurs with
non-negligible probability. Then, with non-negligible probability,
$S(1^n, s_b)$ aborts due to Case-R2 failure in its last simulation
trial with respect to the  covered public-key set $\mathcal{C}_b$,
and the value committed in $\tilde{c}^{(i)}_{crs}$ (in the
successful $i$-th right session w.r.t. an uncovered public-key
$PK^{(j)}_L\not \in \mathcal{C}_b \cup \{PK_L, PK_R\}$ during the
simulation trial w.r.t. $\mathcal{C}_b$)  is the preimage of
$y_b$. Recall that, the successful $i$-th right session is also
the first successful session w.r.t. an uncovered public-key
different from $PK_L$ during the simulation trial w.r.t.
$\mathcal{C}_b$.
 It is easy to see that, with  the same probability, 
  the value committed in
$\tilde{c}^{(i)}_{crs}$ in the $i$-th right successful session
(which is also  the first successful session w.r.t. an uncovered
public-key not in $\mathcal{C}_b \cup \{PK_L, PK_R\}$)  in
$E_2(1^n, s_b, \mathcal{C}_b)$  is  
the preimage of $y_b$. We will use this fact to violate the
one-left-many-right simulation/extraction of commit-then-PRZK with
adaptively setting input and tag for the one left-session, where
the simulator/extractor of commit-then-PRZK first commits  to 0
and then runs the one-left-many-right simulator/extractor of PRZK.

 Before proceeding the analysis, we first present some observations on
 commit-then-PRZK with restricted input selection and
indistinguishable auxiliary information. 
 Consider the following experiments:  $\textup{EXPT}(1^n, w^b, aux^b)$,
 where $w^b\in \{0, 1\}^n$ for $b\in \{0, 1\}$. In $\textup{EXPT}(1^n,  w^b, aux^b)$, the
commit-then-PRZK for $\mathcal{NP}$ is run concurrently,  and a
many-left-many-right CMIM adversary $\mathcal{A}$, possessing
auxiliary information $aux^b$, can set the inputs and tags to
prover instances of left sessions  with the following restriction:
for any $x_i$, $1\leq i\leq s(n)$,
 set by $\mathcal{A}$ for the $i$-th left session of
 commit-then-PRZK, the fixed value $w^b$ is always a valid $\mathcal{NP}$-witness.
 In other words, although $\mathcal{A}$ has the power of adaptive
input selection for provers, but there exists fixed witness-pair
$(w^0, w^1)$ for all inputs selected by $\mathcal{A}$. Such
adversary is called \emph{restricted} input-selecting
CMIM-adversary. Denote by $trans^b$ the transcript of the
experiment $\textup{EXPT}(1^n, w^b, aux^b)$ (i.e., the view of
$\mathcal{A}$ in
 $\textup{EXPT}(1^n, w^b, aux^b)$), and by
$\widehat{W}^b=\{\hat{w}^b_1, \cdots, \hat{w}^b_{s(n)} \}$
 the witnesses encoded (determined)  by the
statistically-binding commitments (at the beginning) of successful
right sessions in $trans^b$; For  a right session  that aborts or
the tag of the underlying PRZK is identical to that  in one of
left sessions,
 $\hat{w}^b_i$ is set to be  a special symbol $\bot$. We
want to show the following proposition: 

\begin{proposition}\label{Case2}
If the ensembles  $\{aux^0\}_{n\in N, w^0\in \{0, 1\}^n, w^1\in
\{0, 1\}^n}$ and $\{aux^1\}_{n\in N, w^0\in\{0, 1\}^n, w^1\in\{0,
1\}^n}$ are indistinguishable, then the ensembles  $\{(trans^0,
\widehat{W}^0)\}_{n\in N, w^0\in \{0, 1\}^n, w^1\in \{0, 1\}^n}$
in accordance with  $\textup{EXPT}(1^n,\\  w^0, aux^0)$   and
$\{(trans^1, \widehat{W}^1)\}_{n\in N, w^0\in\{0, 1\}^n,
w^1\in\{0, 1\}^n}$ in accordance with  $\textup{EXPT}(1^n, w^1,
aux^1)$  are also indistinguishable.
\end{proposition}
\textbf{Proof} (of Proposition  \ref{Case2}): 
This is established by investigating a series of  experiments.


First consider two experiments  $\textup{EXPT}^n_1(1^n,  w^b,
aux^b)$, where $b\in \{0, 1\}$.  In $\textup{EXPT}^n_1(1^n, w^b,
aux^b)$, a \emph{one-left-many-right} restricted input-selecting
MIM adversary $\mathcal{A}$, possessing auxiliary information
$aux^b$, interacts with the prover instance of commit-then-PRZK in
one left session and sets the input $x$ of the left session  such
that $(x, w^b)\in \mathcal{R}_\mathcal{L}$, and concurrently
interacts with many honest verifier instances on the right. From
the one-many simulation/extraction SZKAOK property of PRZK (with
adaptively setting input and tag for the one left session) and
computational-hiding property of the underlying
statistically-binding commitments, by hybrid arguments, we can
conclude that if $aux^0$ is indistinguishable from $aux^1$, then
$\mathcal{A}$'s views and the witnesses encoded (actually
\emph{extracted}) in the two experiments, i.e., $(trans^0,
\widehat{W}^0)$ and $(trans^1, \widehat{W}^1)$), are
indistinguishable. Specifically, consider that the one left
session is simulated by first committing to 0 and then running
the simulator/extractor of PRZK. 

In more details, due to the statistical ZK property of PRZK, for
any bit  $b\in \{0, 1\}$ $(trans^b, \widehat{W}^b)$ in
$\textup{EXPT}^n_1(1^n, w^b, aux^b)$ is identical to $(trans^b,
\widehat{W}^b)$ in a modified version of $\textup{EXPT}^n_1(1^n,
w^b, aux^b)$, called commit($w^b$)-then-simulatedPRZK,  in which
the PRZK of the one left session is simulated rather than really
executed (but the witness $w^b$ is still committed to the
statistically-binding commitment of the left session).
 Then, for this experiment, due to the computational hiding property of the statistically-binding commitment scheme used in
 commit-then-PRZK,
  $(trans^b, \widehat{W}^b)$ of  the
commit($w^b$)-then-simulatedPRZK experiment is computationally
indistinguishable from that of  the commit($0$)-then-simulatedPRZK
experiment in which ``0" (rather than $w^b$) is committed to the
statistically-binding commitment of  the one left session.



Now we consider the following  two experiments:
$\textup{EXPT}(1^n, w, aux^b)$,  where $b\in \{0, 1\}$ and $w\in
\{w^0, w^1\}$. In $\textup{EXPT}(1^n, w, aux^b)$, a
many-left-many-right restricted input-selecting  MIM adversary
$\mathcal{A}$, possessing auxiliary information $aux^b$, interacts
concurrently with many prover instances  on the left (such that
$w$ is always a witness for inputs selected adaptively by
$\mathcal{A}$ for left sessions),  and interacts with many honest
verifier instances on the right. Then, the indistinguishability
between the ensembles $\{(trans^0, \widehat{W}^0)\}_{n\in N,
w^0\in\{0, 1\}^n, w^1\in \{0, 1\}^n}$ and $\{(trans^1,
\widehat{W}^1)\}_{n\in N, w^0\in\{0, 1\}^n, w^1\in \{0, 1\}^n}$ is
direct from the indistinguishability between $\{aux^0 \}_{n\in N,
w^0\in \{0, 1\}^n, w^1\in \{0, 1\}^n}$ and $\{aux^1\}_{n\in N,
w^0\in\{0, 1\}^n, w^1\in \{0, 1\}^n}$ and the adaptive
one-left-many-right simulation-extractability of PRZK.
Specifically, this is derived by a simple reduction  to the above
one-left-many-right case. Note that according to the definition of
indistinguishability between \emph{ensembles}, $(w, aux^0)$ and
$(w, aux^1)$ are indistinguishable. \emph{Actually, $(w^0, w^1,
aux^0)$ and $(w^0, w^1,  aux^1)$ are indistinguishable.} Also,
note that all sessions 
 in $\textup{EXPT}(1^n, w, aux^b)$ can be emulated internally by a PPT
algorithm given $(w, aux^b)$.


We return back to investigate   the  experiments:
$\textup{EXPT}(1^n, w^b, aux^b)$ with respect to
many-left-many-right restricted input-selecting MIM adversary
$\mathcal{A}$. Firstly, the distribution ensemble of  \\
$\{(trans^0, \widehat{W}^0)\}_{n\in N, w^0\in\{0, 1\}^n, w^1\in
\{0, 1\}^n}$ in accordance with  $\textup{EXPT}(1^n, w^0, aux^0)$
and the distribution ensemble of $\{(trans^0,
\widehat{W}^0)\}_{n\in N, w^0\in\{0, 1\}^n, w^1\in \{0, 1\}^n}$
\emph{in accordance with $\textup{EXPT}(1^n, w^0, aux^1)$ } are
indistinguishable,  if $\{aux_0\}_{n\in N, w^0\in\{0, 1\}^n,
w^1\in \{0, 1\}^n}$ and $\{aux_1\}_{n\in N, w^0\in\{0, 1\}^n,
w^1\in \{0, 1\}^n}$ are indistinguishable, where
$\textup{EXPT}(1^n, w^0, aux^1)$ denotes a hybrid experiment in
which the CMIM adversary possesses auxiliary information $aux^1$
while concurrently interacting on the left with many  prover
instances of  the fixed witness $w^0$. Then, by a
simple hybrid argument to the one-left-many-right case, 
 we get that the distribution ensemble  $\{(trans^0, \widehat{W}^0)\}_{n\in N,
w^0\in\{0, 1\}^n, w^1\in \{0, 1\}^n}$ in accordance with
$\textup{EXPT}(1^n, w^0, aux^1)$ is indistinguishable from  the
distribution ensemble of  $\{(trans^1, \widehat{W}^1)\}_{n\in N,
w^0\in\{0, 1\}^n, w^1\in \{0, 1\}^n}$ in accordance with
$\textup{EXPT}(1^n, w^1, aux^1)$. In more detail, if the above
ensembles are distinguishable, then the difference can be reduced,
by hybrid arguments, to the difference of witnesses used in only
one
left session. Note that, 
 all sessions other than the one  left
session  can be emulated internally by a PPT algorithm given
$(w^0, w^1,  aux^1)$.

 Proposition \ref{Case2} follows. \hfill
$\square$

Now, we return back to the experiments $E(1^n, s_b)$ for finishing
the proof of Lemma \ref{R2}. We first prove  that
$\{\mathcal{C}_0\}_{n\in N, s_0\in\{0, 1\}^n, s_1\in \{0, 1\}^n}$
is indistinguishable from
 $\{\mathcal{C}_1\}_{n\in N,
s_0\in\{0, 1\}^n, s_1\in \{0, 1\}^n}$ according to  the analysis
of Proposition \ref{Case2}, where
 $\mathcal{C}_b$, $b\in \{0, 1\}$,  is the set
of extracted-keys
 (corresponding to public-keys in $F-\{PK_R\}$) that is  used by the simulator $S(1^n, s_b)$
  in its last simulation
 repetition. Equivalently,    $\mathcal{C}_b$ is generated by $E_1(1^n, s_b)$ and is passed on to $E_2$.
Note that  $s_b=SK_R$ is the simulated secret-key used by $S$
(equivalently, $E_1$).
 Actually, we can show that for any $k$, $1\leq k\leq s(n)+1$, if the
 distribution ensemble of the set of extracted-keys used in the $(k-1)$-th simulation
 repetition of $S(1^n, s_0)$
 using $SK_R=s_0$,
 denoted $\{\mathcal{C}^{k-1}_0\}_{n\in N,
s_0\in\{0, 1\}^n, s_1\in \{0, 1\}^n}$,  is indistinguishable from
that of $\{\mathcal{C}^{k-1}_1\}_{n\in N, s_0\in\{0, 1\}^n, s_1\in
\{0, 1\}^n}$
 (the set of extracted-keys used in the $(k-1)$-th
 simulation repetition of $S(1^n, s_1)$), then the distribution ensembles of $\{\mathcal{C}^k_0\}_{n\in N,
s_0\in\{0, 1\}^n, s_1\in \{0, 1\}^n}$ and
 $\{\mathcal{C}^k_1\}_{n\in N,
s_0\in\{0, 1\}^n, s_1\in \{0, 1\}^n}$ are also indistinguishable.

 We consider an imaginary simulator $\hat{S}(s_b, \mathcal{C}^{k-1}_b)$, who mimics
 the experiment $E_2(s_b, \mathcal{C}^{k-1}_b)$  with respect to the set of extracted-keys
  $\mathcal{C}^{k-1}_b$.
  We remark that the run of $\hat{S}(1^n,s_b,
 \mathcal{C}^{k-1}_b)$  actually amounts to the experiment
 $\textup{EXPT}(1^n, w^b, aux^b)$ defined in Proposition
 \ref{Case2}, where $s_b$ amounts to $w^b$ and $\mathcal{C}^{k-1}_b$ amounts to
 $aux^b$. Actually, $\hat{S}(1^n,s_b,
 \mathcal{C}^{k-1}_b)$   amounts to a restricted version of
 $\textup{EXPT}(1^n, w^b, aux^b)$. Specifically, $\hat{S}$ (who  incorporates
 $\mathcal{C}^{k-1}_b$ and  internally runs the CMIM adversary
 $\mathcal{A}$) amounts to a many-left-\emph{one}-right adversary
 against commit-then-PRZK, in which it concurrently interacts with its oracle (i.e.,
 the prover of commit-then-PRZK $P(s_b)$)
 in the many left sessions and the only one right session is just the one
 (the successful Stage-1 of a right session or Stage-5 of a left session
 in $\hat{S}(1^n,s_b,
 \mathcal{C}^{k-1}_b)$) in which $\mathcal{A}$ successfully finishes
 the commit-then-proof w.r.t. an uncovered public-key not in
 $\mathcal{C}^{k-1}_b \cup \{PK_L, PK_R\}$. By applying Proposition
 \ref{Case2}, it is easy to see that if the ensembles
 $\{\mathcal{C}^{k-1}_0\}_{n\in N,
s_0\in\{0, 1\}^n, s_1\in \{0, 1\}^n}$ and
$\{\mathcal{C}^{k-1}_1\}_{n\in N, s_0\in\{0, 1\}^n, s_1\in \{0,
1\}^n}$ are
 distinguishable,  $\{\mathcal{C}^{k}_0\}_{n\in N,
s_0\in\{0, 1\}^n, s_1\in \{0, 1\}^n}$ and
$\{\mathcal{C}^{k}_1\}_{n\in N, s_0\in\{0, 1\}^n, s_1\in \{0,
1\}^n}$ are
 also  distinguishable.  Finally, note that $\mathcal{C}^{0}_0$ and $\mathcal{C}^{0}_1$ (the set of
 extracted-keys corresponding to $F-\{PK_R\}$ at the beginning of the simulation)  are
 identical, i.e.,
 both of them are empty set. By inductive steps, we get that  the
 distribution ensembles of $\{\mathcal{C}^{k}_0\}_{n\in N,
s_0\in\{0, 1\}^n, s_1\in \{0, 1\}^n}$ and
$\{\mathcal{C}^{k}_1\}_{n\in N, s_0\in\{0, 1\}^n, s_1\in \{0,
1\}^n}$ are  indistinguishable for any $k$, $1\leq k \leq s(n)+1$.

But,  suppose Case-2 of Case-R2 failure  occurs with
non-negligible
probability. That is, 
with  the same probability, 
  the value committed in
$\tilde{c}^{(i)}_{crs}$ in the successful  $i$-th right session
for some $i$, $1\leq i\leq s(n)$, which  is also the first
successful session w.r.t. an uncovered public-key not in
$\mathcal{C}_b \cup \{PK_L, PK_R\}$, in
$E_2(1^n, s_b, \mathcal{C}_b)$ is  
the preimage of $y_b$.  It can be directly checked that  the tag
used by Stage-5 of the $i$-th right session, $(PK^{(j)}_L,
r^{(i)}_r, \tilde{r}^{(i)})$ where  $PK^{(j)}_L\not\in
\mathcal{C}_b \cup \{PK_L, PK_R\}$ and $r^{(i)}_r$ is a random
$n$-bit string, must be different from the tags used by the prover
$P(1^n, s_b)$ of commit-then-PRZK (i.e., the oracle of $E_2$).
Recall that the tags of Stage-1 of right sessions (\emph{run by
$P(s_b)$})
 is of the form $(\cdot, y_0, y_1)$ and the tags of Stage-5  of left sessions
 (\emph{run by  $P(s_b)$})   is of the form $(PK_L, \cdot, \cdot)$. Also, note that
$E_2$ actually amounts to a many-left-\emph{one}-right CMIM
adversary, that is, all interactions except for the interactions
with the prover $P(s_b)$ of commit-then-PRZK and the Stage-5 of
the successful $i$-th right session can be internally emulated by
$E_2$. This means that, given oracle access to the prover $P(s_b)$
of commit-then-PRZK and  the indistinguishable
$\{\mathcal{C}_b\}_{n\in N, s_0\in\{0, 1\}^n, s_1\in \{0, 1\}^n}$
, $E_2$ can successfully commit the preimage of $y_b$ in the
successful $i$-th right session with different tag, which violates
Proposition \ref{Case2}. This shows that Case-2 of Case-R2 failure
can occur also with negligible probability. Thus, Case-R2 failure
can occur with at most negligible probability. This finishes the
proof of Lemma \ref{R2}, from which the simulatability of the CNM
simulation depicted in Figure \ref{CNMCTsimu} is then established.
\hfill $\square$

\begin{itemize}
\item \textsf{Strategy-restricted and predefinable  randomness}
\end{itemize}

 Now, we proceed to show the strategy-restricted and predefinable
 randomness property of the CNM simulator $S$ depicted in Figure
 \ref{CNMCTsimu}. Denote by $R_L=\{R^{(1)}_L, R^{(2)}_L, \cdots
 R^{(s(n))}_L\}$ the coin-tossing outputs of the $s(n)$ left
 sessions in $str$ (i.e., the first output of $S$), and by $sta_L=\{sta^{(1)}_L, sta^{(2)}_L,
  \cdots, sta^{(s(n))}_L\}$ the state information corresponding to $R_L$ included in $sta$
  (i.e., the second output of $S$). Similarly, denote by  $R_R=\{R^{(1)}_R, R^{(2)}_R, \cdots
 R^{(s(n))}_R\}$ the coin-tossing outputs of the $s(n)$ right
 sessions in $str$, and by  $sta_L=\{sta^{(1)}_R, sta^{(2)}_R,
  \cdots, sta^{(s(n))}_R\}$ the state information for $R_R$. We
  want to show that, with overwhelming probability, the distributions of $(R_L, sta_L)$ and
$(R_R, sta_R)$ are identical to that of $\mathcal{M}^{s(n)}_{CRS}
(1^n)$. Recall that,  $(\{r_1, r_2, \cdot, r_{s(n)}\},
\{\tau_{r_1}, \tau_{r_2}, \cdots, \tau_{r_{s(n)}}\})
\longleftarrow \mathcal{M}^{s(n)}_{CRS}(1^n)$ denotes  the output
of the experiment of running $\mathcal{M}_{CRS}(1^n)$
\emph{independently} $s(n)$ times.

Note that, according to the CNM simulation described in Figure
\ref{CNMCTsimu}, for any $i$, $1\leq i\leq s(n)$,  the output of
the $i$-th left session, i.e., $R^{(i)}_L$,  in the simulation is
always $S_L^{(i)}$ and $sta^{(i)}_L$ is always $\tau^{(i)}_{L}$,
where $(S^{(i)}_L, \tau^{(i)}_L)$ is the output of an independent
run of $\mathcal{M}_{CRS}(1^n)$. It is directly followed that the
distribution of $(R_L, sta_L)$ is identical to that of
$\mathcal{M}^{s(n)}_{CRS} (1^n)$.

The complicated point here is to show that, with overwhelming
probability,  the distribution of
 $(R_R, sta_R)$ is also  identical to that of $\mathcal{M}^{s(n)}_{CRS}
(1^n)$. According to the CNM simulation depicted in Figure
\ref{CNMCTsimu}, if we can prove that, with overwhelming
probability, for any $i$, $1\leq i\leq s(n)$, the coin-tossing
output of the successful $i$-th right session $R^{(i)}_R$ is
either $S^{(i)}_R$ or $R^{(k)}_L=S^{(k)}_L$ for some $k$, $1\leq
k\leq s(n)$;  furthermore,  any left-session output $S^{(k)}_L$
can be the coin-tossing output for    at most \emph{one}
successful right session (which implies the coin-tossing outputs
of successful right sessions are independent), then the
distribution of $(R_R, sta_R)$ is also identical to that of
$\mathcal{M}^{s(n)}_{CRS} (1^n)$.

For any $i$, $1\leq i\leq s(n)$, we consider the successful $i$-th
right session with respect to a public-key $PK^{(j)}_L$. As we
have shown that Case-R2 failure occurs with negligible
probability, we get $PK^{(j)}_L \in \mathcal{C}_b \cup \{PK_R,
PK_L\}$, where $\mathcal{C}_b$ is the set of extracted-keys
(corresponding to public-keys in $F-\{PK_R\}$)  used by $S(s_b)$
in its last simulation repetition.

We first observe that, if $PK^{(j)}_L=PK_L$ then with overwhelming probability  the tag of
Stage-5 of  the successful $i$-th right session must be identical
to that of Stage-5 of a left session simulated by the simulator
$S$. Recall that  the  all Stage-5 tags of  right sessions are
different strings, as they contain random Stage-3 strings sent by
the simulator. This means that Stage-5 tags of right sessions are
also different from Stage-1 tags of right sessions simulated by
$S$ (note that all Stage-1 tags of right
sessions consist of the fixed $PR_R$).  
 Now, suppose the Stage-5 tag of the successful $i$-th right session
is also different from the Stage-5 tags
 of all left sessions simulated by $S$, then it implies that the tag used by the CMIM adversary
for Stage-5 of the $i$-th right session is different from all tags
used by the simulator (equivalently, the prover $P(s_b)$ of
commit-then-PRZK in the experiment $E$ in the analysis of Lemma
\ref{R2}). By the AOK property, it implies that the value
committed to  $\tilde{c}^{(i)}_{crs}$ (sent by $\mathcal{A}$ in
Stage-5 of the $i$-th right session) can be extracted. We consider
the possibilities of the value committed to
$\tilde{c}^{(i)}_{crs}$:
\begin{itemize}
\item  By the one-wayness of $y_{1-b}$ the value committed cannot
be the preimage of $y_{1-b}$; \item  According to the analysis of
Lemma \ref{R2}, the value also cannot be the preimage of $y_b$.
\end{itemize}
 Thus, the value committed (that can be extracted) will
be the secret-key of $PK_L$, which however violates the
one-wayness of $PK_L$ as the simulator never knows and uses the
secret-key of $PK_L$ in its simulation. Thus, we conclude that, if
a successful right session is w.r.t. $PK_L$, the tag used by
$\mathcal{A}$ for commit-then-PRZK of Stage-5  must be identical
to that of one left-session simulated by $S$. As the Stage-5 tag
consists of the coin-tossing output, i.e., the Stage-4 message,
this means that the coin-tossing output of the $i$-th right
session must  be $R^{(k)}_L=S^{(k)}_L$ for some $k$, $1\leq k\leq
s(n)$.

Now, we consider the case $PK^{(j)}_L \neq PK_L$. In this case,
$S$ has already learnt the corresponding secret-key $SK^{(j)}_L$.
Now, suppose the coin-tossing output of the successful $i$-th
right session is neither  $S^{(i)}_R$ nor $R^{(k)}_L=S^{(k)}_L$
for all $k$, $1\leq k\leq s(n)$. This implies that the Stage-5 tag
used by $\mathcal{A}$ in the successful $i$-th right session is
different from Stage-5 tags of all left sessions \footnote{Note
that all Stage-5 tags of left sessions are of the form $(PK_L,
\cdot, \cdot)$, and the Stage-5 tag of the successful  $i$-th
right session is of the form $(PK^{(j)}_L, \cdot, \cdot)$ for
$PK^{(j)}_L \neq PK_L$. } as well as the Stage-1 tags of all right
sessions simulated by $S$. Again, by the AOK property, we consider
the value committed to $\tilde{c}^{(i)}_{crs}$: According to the
simulation of $S$, it always sets Stage-3 message $r^{(i)}_r$ of
right session to be
$PRF_{SK^{(j)}_L}(\tilde{r}^{(i)\prime}_l)\oplus S^{(i)}_R$, where
$\tilde{r}^{(i)\prime}_l$ is the Stage-2 message of the $i$-th
right session sent by the CMIM adversary $\mathcal{A}$. Suppose
the coin-tossing output of the successful $i$-th right session is
not $S^{(i)}_R$, then the value committed to
$\tilde{c}^{(i)}_{crs}$ cannot be $SK^{(j)}_L$, which will be the
preimage of either $y_{1-b}$ or $y_b$. But, each case reaches the
contradiction: committing to the preimage of $y_{1-b}$ is
impossible due to the one-wayness of $y_{1-b}$; committing to the
preimage of $y_b$ violates the one-left-many-right
non-malleability of PRZK as demonstrated in the analysis of Lemma
\ref{R2}.  So, we conclude that, with overwhelming probability,
for any successful right session the coin-tossing output is either
the independent value $S^{(i)}_R$ or $S^{(k)}_L$ for some $k$,
$1\leq k\leq s(n)$ (i.e., the coin-tossing output of one left
session).

To finally establish the property of      strategy-restricted and
predefinable randomness, we need to further show, for any
$S^{(k)}_L$ it can occur as Stage-4 message (i.e., the
coin-tossing output) for  \emph{at most one successful right
session}. Suppose there are $i_0, i_1$, $1\leq i_0\neq i_1 \leq
s(n)$, such that both of the $i_0$-th right session and the
$i_1$-th right session are successful with the same Stage-4
message $S^{(k)}_L$. Recall that the Stage-5 tag  of each of the
two right sessions includes the same $S^{(k)}_L$ as well as a
random Stage-3 message sent by the simulator; Also note that the
$S^{(k)}_L$ can appear, as a part of Stage-5 tag as well as
coin-tossing output, for at most one left session (all
coin-tossing outputs, i.e., Stage-4 messages,  of left sessions
are independent random strings output by $\mathcal{M}_{CRS}$).
This implies that, there must exist a bit $b$ such that the
Stage-5 tag of the $i_b$-th right session is different from all
Stage-5 tags of left sessions (run by the simulator) and Stage-1
tags of right sessions (run by the simulator). According to above
clarifications and analysis, with overwhelming probability, the
(left-player) public-key $PK^{(j)}_L$
 used by $\mathcal{A}$ in the $i_b$-th successful right
session is \emph{covered} and is  \emph{not} $PK_L$, and the value
committed in $\tilde{c}^{(i_b)}_{crs}$ is neither the secret-key
of
the \emph{covered} public-key $PK^{(j)}_L$ 
 nor the preimage of $y_{1-b}$; Also,  the value committed  cannot
be the preimage of $y_{b}$ in accordance with the analysis of
Lemma \ref{R2}. Contradiction is reached in either case.
 \hfill $\square$


\begin{itemize}
\item \textsf{Secret-key independence}
\end{itemize}

Specifically, we need to show that $\Pr[\mathcal{R}(SK_R, str,
sta)=1]$ is negligibly close to $\Pr[\mathcal{R}(SK^{\prime}_R, str, sta)\\
=1]$ for any polynomial-time computable relation $\mathcal{R}$.
 In more details, for any pair $(s_0, s_1)$ in the (simulated right-player) key-generation stage,
denote by $(str^b, sta^b)$ the output of $S(1^n, s_b)$ when it is
using $SK_R=s_b$. Then,   $\Pr[\mathcal{R}(SK, str, sta)=1]
=\frac{1}{2}\Pr[\mathcal{R}(s_0, str^0, sta^0)=1| S \ \text{uses}
\
  SK_R=s_0\
\text{in} \  \text{generating}\   (str^0,  sta^0)]+
\frac{1}{2}\Pr[\mathcal{R}(s_1, str^1, sta^1)=1|S\ \text{uses}\
SK_R=s_1 \ \text{in generating } (str^1, sta^1)]$, and
$\Pr[\mathcal{R}(SK^{\prime}_R, str, sta)=1]
=\frac{1}{2}\Pr[\mathcal{R}(s_0, str^1, sta^1)=1| S \ \text{uses}
\ SK_R=s_1 \text{in}\ \text{generating}\  (str^1, sta^1)]+
\frac{1}{2}\Pr[\mathcal{R}(s_1, str^0, sta^0)=1|S\ \text{uses}\
SK_R=s_0 \ \text{in generating } (str^0, sta^0)]$.
 Suppose the  secret-key independence
property does not hold, it implies that there exists a bit $\alpha
\in \{0, 1\}$ such that the difference between
$\Pr[\mathcal{R}(s_\alpha, str^0, sta^0)=1|S \ \text{uses} \ s_0 \
\text{in} \
 \text{generating } (str^0, sta^0)]$ and $\Pr[\mathcal{R}(s_\alpha, str^1, sta^1)=1|S \
\text{uses} \ s_1 \ \text{in}\   \text{generating } (str^1,
sta^1)]$ is non-negligible. It implies that $(s_\alpha, str^0,
sta^0)$
  and $(s_\alpha, str^1, sta^1)$
  are distinguishable.
  But, note
that the analysis of Lemma \ref{R2} and Proposition \ref{Case2}
has already established that the distribution  ensembles of
$\{S(1^n, s_0)=(str^0, sta^0)\}_{n\in N, s_0\in\{0, 1\}^n, s_1\in
\{0, 1\}^n}$ and $\{S(1^n, s_1)=(str^1, sta^1)\}_{n\in N,
s_0\in\{0, 1\}^n, s_1\in \{0, 1\}^n}$ are indistinguishable.
Specifically, the distribution ensembles of the sets of
extracted-keys corresponding to the  public-keys in $F-\{PK_R\}$,
$\{\mathcal{C}_0\}_{n\in N, s_0\in\{0, 1\}^n, s_1\in \{0, 1\}^n}$
and $\{\mathcal{C}_1\}_{n\in N, s_0\in\{0, 1\}^n, s_1\in \{0,
1\}^n}$ used by $S(1^n, s_b)$ for $b\in \{0, 1\}$ in the last
simulation repetition,  are indistinguishable, and then the
indistinguishability between the ensembles  $\{(str^0,
sta^0)\}_{n\in N, s_0\in\{0, 1\}^n, s_1\in \{0, 1\}^n}$
and $\{(str^1, sta^1)\}_{n\in N, s_0\in\{0, 1\}^n,\\
s_1\in \{0, 1\}^n}$ are from  Proposition \ref{Case2}.



The proof of Theorem \ref{CNMCTtheo} is finished. \hfill $\square$


\vspace{0.5cm}

\noindent\textbf{Acknowledgments.} We are  much indebted  to Fraces F. Yao for her many valuable discussions and contributions to this work (though she declined the coauthorship of this work).   The third author thanks Rafael Pass and  Alon
Rosen for helpful  discussions. 





\end{document}